\documentclass[]{aastex631}
\shorttitle{AASTeX v6.3.1 Sample article}
\shortauthors{Deesamutara et al.}
\graphicspath{{./}{figures/}}

\usepackage{color}                 
\usepackage{amsmath}               
\usepackage{caption} 
\usepackage{subcaption}             
\usepackage[para,online,flushleft]{threeparttable} 
\usepackage{multirow}               
\usepackage{float}

\begin{document}
\title{Extracting the X-ray reverberation response functions from the AGN light curves using an autoencoder}

\correspondingauthor{Poemwai Chainakun}
\email{pchainakun@g.sut.ac.th}

\author[0000-0002-0964-0050]{Sanhanat Deesamutara}
\affiliation{School of Physics, Institute of Science, Suranaree University of Technology, Nakhon Ratchasima 30000, Thailand}

\author[0000-0002-9099-4613]{Poemwai Chainakun}
\affiliation{School of Physics, Institute of Science, Suranaree University of Technology, Nakhon Ratchasima 30000, Thailand}
\affiliation{Center of Excellence in High Energy Physics and Astrophysics, Suranaree University of Technology, Nakhon Ratchasima 30000, Thailand}

\author{Tirawut Worrakitpoonpon}
\affiliation{School of Physics, Institute of Science, Suranaree University of Technology, Nakhon Ratchasima 30000, Thailand}
\affiliation{Center of Excellence in High Energy Physics and Astrophysics, Suranaree University of Technology, Nakhon Ratchasima 30000, Thailand}

\author{Kamonwan Khanthasombat}
\affiliation{School of Physics, Institute of Science, Suranaree University of Technology, Nakhon Ratchasima 30000, Thailand}

\author[0000-0002-4516-6042]{Wasutep Luangtip}
\affiliation{Department of Physics, Faculty of Science, Srinakharinwirot University, Bangkok 10110, Thailand}
\affiliation{National Astronomical Research Institute of Thailand, Chiang Mai 50180, Thailand}

\author[0000-0002-9639-4352]{Jiachen Jiang}
\affiliation{Department of Physics, University of Warwick, Gibbet Hill Road, Coventry CV4 7AL, UK}
\affiliation{Institute of Astronomy, University of Cambridge, Madingley Road, Cambridge CB3 0HA, UK}

\author[0000-0002-6716-4179]{Francisco Pozo Nuñez}
\affiliation{Astroinformatics, Heidelberg Institute for Theoretical Studies, Schloss-Wolfsbrunnenweg 35, D-69118 Heidelberg, Germany}

\author[0000-0003-3626-9151]{Andrew J. Young}
\affiliation{H.H. Wills Physics Laboratory, Tyndall Avenue, Bristol BS8 1TL, UK}

\begin{abstract}
We study the X-ray reverberation in active galactic nuclei (AGN) using the variational autoencoder (VAE), which is a machine-learning algorithm widely used for signal processing and feature reconstruction. While the X-ray reverberation signatures that contain the information of the accretion disk and the X-ray emitting corona are commonly analyzed in the Fourier domain, this work aims to extract the reverberation response functions directly from the AGN light curves. The VAE is trained using the simulated light curves that contain the primary X-rays from the lamp-post corona varying its height and the corresponding reflection X-rays from the disk. We use progressively more realistic light-curve models, such as those that include the effects of disk-propagating fluctuations and random noises, to assess the ability of the VAE to reconstruct the response profiles. Interestingly, the VAE can recognize the reverberation patterns on the light curves, hence the coronal height can be predicted. We then deploy the VAE model to the {\it XMM-Newton} data of IRAS~13224--3809 and directly estimate, for the first time, the response functions of this source in various observations. The result reveals the corona changing its height between $3~r_{\rm g}$ and $20~r_{\rm g}$, which is correlated with the source luminosity and in line with previous literature. Finally, we discuss the advantages and limitations of this method.
\\
\\
\noindent{\it Unified Astronomy Thesaurus concepts:} Reverberation mapping (2019); X-ray astronomy (1810); Active galactic nuclei (16); Black hole physics (159)

\end{abstract}


\section{Introduction}

Active galactic nuclei (AGN) are the most luminous objects in the Universe, which are powered by an accretion of gas onto the central supermassive black hole. The accreting gas can form a disk-like structure referred to as an accretion disk that converts the gravitational energy into thermal optical/UV radiation. The disk photons gain more energy by Compton up-scattering with hot electrons inside the corona, raising the photon energy to X-rays \citep{Reynolds2003}. The X-rays from the corona can travel back to the disk, be re-processed, and reflect off the disk in the form of a reflection spectrum, carrying out the characteristic features of the emission lines to an observer \citep{George1991, Ross1999, Ross2005, Garcia2010, Garcia2013}. How much the Doppler and relativistic effects distort the lines determines how close the inner edge of the disk is to the black hole, hence allowing us to probe the innermost region of the AGN. 

The X-ray variations in AGN occur on a wide range of timescales \citep[e.g.][]{Vaughan2003}. The variabilities on long timescales relate to the fluctuations of the mass accretion rate that propagate inwards along the disk, modulating the outer (spectrally softer) regions first before the inner (spectrally harder) regions, hence producing the hard lags \citep{Kotov2001, Arevalo2006}. On shorter timescales, the fluctuations in the direct continuum cause the reflection to reverberate, so the changes in the reflection-dominated energy bands are expected to lag behind the changes in the continuum-dominated band. The amplitude of the reverberation lags relate to the light travel time between the corona and the reflecting region, providing us insights about the disk-corona geometry \citep[see][for a review]{Uttley2014, Cackett2021}. 

Theoretical works in calculating the X-ray reverberation lags have been done early on \citep{Reynolds1999, Young2000} before the first hints of the reverberation were seen in Ark 564 \citep{McHardy2007}, followed by the first robust detection in 1H0707--495 \citep{Fabian2009} where the fluctuations in the soft (reflection dominated) energy band lagged behind those in the hard (continuum dominated) band by $\sim 30$ s. Since then, the number of AGN that exhibit the X-ray reverberation lags have increased \citep[e.g.][]{DeMarco2013, Kara2016}. While the nature of the corona is still uncertain, the dynamics of the corona were evident, e.g., in Mrk~335 \citep{Wilkins2015}, IRAS~13224--3809 \citep{Alston2020, Caballero2020, Chainakun2022b}, 1H~0707--495 \citep{Hancock2023, Mankatwit2023}, and NGC~4051 \citep{Kumari2023}. By studying the time delays between UV and X-ray bands, a change in the geometry of the system (e.g. the height of the corona) was also suggested \citep{Kumari2024}.

Meanwhile, the full-relativistic X-ray reverberation models were continuously developed based on the ray-tracing simulations that trace photon trajectories along the Kerr geodesics between the corona, the disk and the observer \citep[e.g.,][]{Wilkins2013, Cackett2014, Emmanoulopoulos2014, Chainakun2016, Epitropakis2016}. In this way, we can calculate the flux of the reflection photons from different parts of the disk as seen by the observer and as a function of time after the direct continuum. This is known as the response function that can be used to generate realistic light curves based on the convolution theorem. The lags between the light curves in different energy bands then are estimated in the Fourier-frequency space \citep{Uttley2014}. Comparing the lags between the observations and the models is traditionally performed in the Fourier domain, either by fitting the lag-frequency spectra \citep[e.g.][]{Alston2020, Caballero2018, Caballero2020} or the power spectral density (PSD) obtained from the modulus square of the Fourier transform \citep[e.g.][]{Emmanoulopoulos2016, Chainakun2022b}. 

This work aims to develop machine learning (ML) models to extract the response functions from the AGN light curves so that they can be directly fitted with the theoretical response profiles. We employ the variational autoencoder (VAE) which is based on the multilayer perceptron (MLP) for signal reconstruction and reverberation feature extraction. Several applications of autoencoding architectures have been applied to astronomical problems such as SED feature extraction~\citep{frontera2017}, strong lensing identification~\citep{cheng_identifying_2020}, and X-ray binary pattern recognition~\citep{OrwatKapola2021}. However, there was no previous work that has been applied to predict the X-ray reverberation response function. Thus, we are motivated by these previous studies to adopt the VAE to our problem and examine it in terms of performance and sensitivity.

Simulated AGN light curves in several scenarios are produced to test the effectiveness of the VAE model in reconstructing the response functions. This is very challenging since the AGN light curves are known to be stochastic (i.e. different realizations of the same underlying physical process may look entirely different). This work can provide an independent tool to discover the underlying process straightforwardly in the time domain and to constrain the key parameters of the system. In addition, we also apply the model to IRAS~13224--3809 which is the well-known AGN that exhibits strong and variable signatures of the X-ray reverberation due to its dynamic corona, where the coronal height seems to increase with the source luminosity \citep[e.g.][]{Alston2020, Caballero2020, Chainakun2022b}. In brief, IRAS~13224--3809 has been identified as a Narrow-line Seyfert 1 galaxy \citep{Veron-Cetty2006}, located at redshift z of 0.06580;\footnote{Fiducial redshift obtained from \url{http://ned.ipac.caltech.edu}} this is corresponding to the luminosity distance $D_{\rm L}$ of 288 Mpc -- in which the cosmological parameters of $H_{\rm 0}$ = 73 km s$^{-1}$ Mpc$^{-1}$, $\Omega_{\rm matter}$ = 0.27, and $\Omega_{\rm vacuum}$ = 0.73 have been assumed \citep{Jiang2018}. The source was well observed by {\it XMM-Newton} observatory \citep{Jansen2001} mainly in 2011 and 2016 in which its total exposure time is $\sim$ 2000 ks, making this a good observational template for this work. The ultimate goal of this work is to directly derive the response functions of IRAS~13224--3809 and discuss the potential use of this method.

The IRAS~13224--3809 observations and detailed data reduction are given in Section 2. The {\sc kynxilrev} model to produce the disc response functions is explained in Section 3. All light curve models investigated here are presented in Section 4. The ML models to extract the response profiles from the light curves as well as the signal similarity indicators are presented in Section 5. The performance of the models and the results when applying to IRAS~13224--3809 are presented in Section 6. The discussion and conclusion is given in Section 7. 

\section{Observations and data reduction}

To test our proposed method with real observational data, in this work, we obtained the data of IRAS~13224--3809 previously observed by {\it XMM-Newton} observatory from {\it XMM-Newton} Science Archive\footnote{\url{http://nxsa.esac.esa.int}}; they are listed in Table~\ref{tab:observations}. The observational data were reprocessed using the science analysis software (SAS) version 21.0.0\footnote{\url{https://www.cosmos.esa.int/web/xmm-newton/sas}} with the up-to-date set of calibration files in April 2024; the SAS task {\sc epproc} with its default parameter values was used to generate the calibrated event lists for all observations; note that, here, only datasets from pn detector were used in order to obtain good quality timing data. In addition, we also removed the exposure periods that were affected by high background flaring events from the reprocessed event files; this was done manually by inspecting the instrument count rate in the high energy band (10 -- 12 keV) and removing the observational periods with having anomalously high count rate of $>$ 0.4 count s$^{-1}$. This results in a factor of $\sim$20 per cent reduction of observational time of each individual dataset, and a random building of the time gaps in the observational data, typically with the periods of $\sim$0.1 -- 10 ks. The total, remaining exposure time (i.e. a summation of good time intervals; GTIs) of each observation is shown in column 3 of Table~\ref{tab:observations}.

We then extracted background-subtracted light curves of the source with the temporal resolution of 1 second from each individual observation from the events which were flagged as \#XMMEA\_EP and PATTERN $\le$ 4; the source extracting area was defined as a circular region with a radius of 20$\arcsec$ centred on the AGN position while that of the background was 50$\arcsec$ radius circular region located in the source-free area close to and also on the same chip with the source's region; these criteria were applied to create light curves in 0.3--1 keV (reflection dominated) band for all observations.

Finally, we also estimate the quality of the obtained light curves by calculating their signal-to-noise ratio (SNR). We use the definition of the SNR similar to what would be employed when we add noises to the simulated light curves, which can be written as 

\citep{SNR_generator}
\begin{equation}
    \text{SNR} =  \frac{\|\mathbf{s}\|^2_2}{\alpha^2 \|\mathbf{n}\|^2_2}  \;,
    \label{eq:s/n}
\end{equation}
where $\mathbf{s}$ and $\mathbf{n}$ represent the signal and noise, respectively, and $\| \cdot \|^2_2$ denotes the squared $l^2$-norm of the parameters. This equation assumes that noise is composed of zero-mean, unit-variant Gaussian noise, while coefficient $\alpha$ amplifies the noise to the designated level of signal-to-noise ratio. Here, $\alpha$ was fixed at unity in the calculation while the SNR was calculated from the signal and noise powers, i.e. the power spectral density, in the frequency range of 0.0625 -- 1 mHz. This is to enable us to be able to compare the quality of real observational data with our simulated data. The SNR result was presented in column 5 of Table~\ref{tab:observations}. We note that while the noise defined in Equation~\ref{eq:s/n} could represent stochastic noise produced by any processes relating to the source, indeed, the noise detected in observational data might be dominated by -- or at least some fraction could be contributed from -- non-stochastic noise, in particular the detector noise. However, this kind of noise should be ignoreable in our case as it would affect only on pn data below 250 eV.\footnote{\url{https://xmm-tools.cosmos.esa.int/external/xmm_user_support/documentation/uhb/epicdetbkgd.html}; see also \citet{Carter2007}} Furthermore, the observational data used in this work also have good quality, in which their SNRs measured by the traditional method are $\ga$100 (i.e. table 1 of \citealt{Nakhonthong2024}), suggesting that the noise is unlikely to dominate the total signal. Examples of some observed and simulated light curves are presented in Appendix~\ref{app:SNR}.

\begin{table*}
\begin{center}
   \caption{\emph{XMM-Newton} dataset of IRAS~13224--3809 used in this work.} 
   \label{tab:observations}
   \begin{threeparttable}
    \begin{tabular}{lcccc}
    \hline
    Observation ID & ~~~~~Observed date~~~~~ & ~Exposure time$^{a}$~ & $L_{\rm X}$$^{b}$~ & SNR$^{c}$ \\
    & & (ks) & ($\times$ 10$^{42}$erg s$^{-1}$) & \\
    \hline
0673580101	&	2011-07-19	&	50.26	& 4.50 & 26.81\\
0673580201	&	2011-07-21	&	64.63	& 3.72 & 50.06\\
0673580301	&	2011-07-25	&	67.67	& 2.59 & 50.75\\
0673580401	&	2011-07-29	&	98.81	& 6.10 & 62.23\\
0780561301	&	2016-07-10	&	121.21	& 3.42 & 55.46\\
0780561501	&	2016-07-20	&	113.82	& 2.79 & 70.17\\
0780561601	&	2016-07-22	&	114.98	& 5.38 & 112.06\\
0780561701	&	2016-07-24	&	112.39	& 3.14 & 66.60\\
0792180101	&	2016-07-26	&	119.64	& 2.58 & 121.11\\
0792180201	&	2016-07-30	&	118.99	& 3.55 & 72.62\\
0792180301	&	2016-08-01	&	103.28	& 2.56 & 79.20\\
0792180401	&	2016-08-03	&	104.90	& 6.86 & 130.71\\
0792180501	&	2016-08-07	&	111.18	& 4.23 & 62.44\\
0792180601	&	2016-08-09	&	114.51	& 5.92 & 93.38\\
    \hline
    \end{tabular}
    \begin{tablenotes}
    \item \textit{Note:} $^{a}$The useful exposure time after the removal of high background flaring periods. $^{b}$The source X-ray luminosity in 2 -- 10 keV band; the data was obtained from the extended data fig. 2 of \citet{Alston2020}. $^{c}$The Signal to Noise Ratio (SNR) of dataset determined using Equation~\ref{eq:s/n} (see text for further detail). 
    \end{tablenotes}
    \end{threeparttable}
    \end{center}
\end{table*}

\section{Response functions from {\sc kynxilrev}}
 
The {\sc kynxilrev} model is a computational code for X-ray reverberation from the accretion disk around the central black hole \citep{Caballero2018,Dovciak2004a,Dovciak2004b}. It assumes that the disk is geometrically thin, optically thick and extended from the innermost stable circular orbit to 1000~$r_{\rm g}$ ($1\; r_{\rm g} = 1 \; GM/c^2$, where $G$ is the gravitational constant, $M$ is the central mass, and $c$ is the speed of light). In {\sc kynxilrev}, the lamp-post corona is employed, and the {\sc xillver} model \citep{Garcia2010,Garcia2013} with the {\sc XSTAR} code \citep{Kallman2001} is used to calculate the reprocessing on the disk as it is irradiated with an isotropic flash of X-ray continuum (i.e., the primary X-rays) from the corona. The primary X-rays is characterized by a power law with the photon index ($\Gamma$) and with the high energy cut-off. The disk also has a constant density, and its ionization state varies accordingly with the incident flux.

By using {\sc kynxilrev}, the response function of the disk reflection is obtained from the ray-tracing simulations that trace photon paths along the Kerr geodesics between the source, the disk, and the observer. Since we aim to deploy the developed VAE model to IRAS~13224--3809, we fix the central mass, the inclination, and the black hole spin to the values found in previous studies, which are $M=2 \times 10^{6} M_{\odot}$ \citep{Alston2020}, $i=45^{\circ}$ \citep{Caballero2020}, and $a=0.998$ \citep{Jiang2018}. Note that $\Gamma$ may vary among different observations, but to avoid the model degeneracy, we fix $\Gamma =2$. In this work, we use \texttt{xillverD-5.fits} as a pre-calculated table model for {\sc xillver}. Different choices of the table models can lead to different amplitudes of the obtained response functions. Any unspecified {\sc kynxilrev} parameters are set to their default values.

We choose to normalize the area under the response function and use the reflection fraction ($R$) as a normalization parameter to regulate the importance between the X-ray reflection and direct continuum components \citep[e.g.][]{Chainakun2022b}. While a lamp-post assumption fixes the ratio of photons reaching the disc to those reaching infinity, we define $R$ as the ratio of reflection flux to continuum flux. This allows $R$ to vary even with fixed geometry, due to factors such as disc density and specific reflection models used (e.g., {\sc reflionx} and {\sc xillver}). Allowing $R$ to vary produces dilution effects similar to when we vary, e.g., disc density or its ionization state (i.e. the first response time of the response function remains the same while the dilution or normalization changes). Therefore, instead of adjusting a large number of parameters in {\sc kynxilrev} that influence dilution effects, we have chosen to fix all of these and vary only $R$. This allows for some flexibility in our model while still maintaining the core structure of the lamp-post geometry.

\section{Modelled light curves}

The AGN light curve in each energy band consists of the X-ray continuum and the disk reflection components. We investigate the progressively more realistic light-curve models when the effects of the disk-propagating fluctuations are excluded and included, referred to as the REV model and REV+PROP model, respectively. Then, we investigated the case where the random noise is included (REV+PROP+NG model), and where the Poisson noise is also taken into account (REV+PROP+NG+NP model). All models focus on the light curves in the soft (reflection dominated, 0.3--1 keV) energy band given by $s(t)$.

\subsection{The REV model}
Let us assume a driving signal $a(t)$, so that the soft-band light curves can be written as \citep{Emmanoulopoulos2014, Chainakun2016,Epitropakis2016}
\begin{equation}
s(t) = b a(t) + R a(t) \otimes \psi(t) \;, \label{eq:s-band}
\end{equation}

where $a(t) \otimes \psi(t) = \int_0^t a(t^\prime) \psi(t-t^\prime)dt^\prime$. The driving signal $a(t)$ in Equations~\ref{eq:s-band} is generated using the {\tt stingray.simulator} from the power spectral density (PSD) in the form of a power-law red noise \citep{Huppenkothen2019}. The slope of the PSD is randomly varied between 1 and 3, covering the mean index of $\sim 2$ as reported in \cite{Gonzalez2012}. The parameter $b$ is the normalization factors tied to the continuum flux contribution in this band:
\begin{equation}
  b  = \int_{0.3~{\rm keV}}^{1~{\rm keV}} E^{-\Gamma} {\rm d} E \;.
    \label{eq:b-norm}
\end{equation}

$\psi(t)$ is the disk response function due to a lamp-post source and $R$ is the normalization related to the reflection fraction. The first and the second terms on the right-hand sides of the light curve equation then represent the flux contributed by the direct continuum and the X-ray reflection, respectively.

Furthermore, the area under $\psi(t)$ is normalized to $b$, so that $R=1$ means an equal contribution between the continuum and the reflected flux in the corresponding energy band. Given the complexity of the model and to avoid a prohibitively large number of parameters, the photon index of the X-ray continuum is fixed at $\Gamma = 2$ in this model. 

We produce the model grid of $\psi(t)$, for an ensemble of 29 coronal heights: [$h=2.3,3,4,5,6,...,30~r_{\rm g}$]. For each grid of $h$, we change the seed in the light curve simulator in {\tt stingray.simulator} to produce 100 different $a(t)$, while uniformly randomly select the reflection fraction in the soft band within the range of $R \in [1,3]$. Therefore, the data contain 2,900 light curves in total. We then investigate the performance of the models to extract the response functions from these light curves. This represents the simplest model with which we begin our investigation. In the final model applied to the IRAS~13224--3809 data, the range of $R$ is extended to $[1, 10]$, and additional factors such as disc-propagating fluctuations and noise effects are incorporated.

\subsection{The REV+PROP model}
This model is built by including additional top-hat response functions, ${\rm TH}(t)$, to the light curve equation. The top-hat function represents the response due to the disk-propagating fluctuations that produce the hard lags on longer timescales than the soft reverberation lags \citep[e.g.][]{Lyubarskii1997,Kotov2001,Arevalo2006}. The shape of the top-hat response is controlled by the centroid, $\tau$, and the width, $w$. In this case, we generalize the light curve equation so that it can represent the light curve in any energy bands of interest. The area under both $\psi(t)$ and ${\rm TH}(t)$ is normalized to 1. The light curve formula is given by \citep[e.g.][]{Alston2014, Chainakun2023}
\begin{equation}
 s(t) = a(t) + R \; a(t) \otimes \psi(t) + P \; a(t) \otimes {\rm TH}(t) \;, \label{eq:s-rev+prop}
\end{equation}
where the normalization $R$ and $P$ regulate the importance of the responses due to the reverberation and the disk propagating-fluctuation processes, respectively. A uniform distribution function is used to generate a random number for $R \in [1,3]$ and $P \in [1,3]$. When $R > P$, the light curve is dominated by the effects of the X-ray reverberation, and when $P > R$, the propagating fluctuations dominate. We note that this study does not account for the potential contribution of a warm corona to the X-ray soft excess \citep[e.g.][]{Kubota2018, Middei2018, Porquet2018,  Ursini2020, Xu2021, Ballantyne2024}. Incorporating this component may affect the dilution of time lags that play a similar role as to adjust the reflection fraction. See Appendix~\ref{app:perfeval} for more discussion in the case that $R$ of the real data is different from that of the training data.
 
We produce the model grids for $\psi(t)$ similar to what is done in the REV model. On the other hand, the ${\rm TH}(t)$ is produced by the uniform distribution function, where the centroid and the width are randomly selected from $\tau \in [10^{4},5 \times 10^{4}]$~s and $w \in [10^{3},5 \times 10^{3}]$~s. This ensures that the average response time and the width of the response of ${\rm TH}(t)$ are larger than those of $\psi(t)$, which is likely the case for the disk propagating-fluctuation lags that dominate on longer timescales. In this case, we simulate 2,900 light curves using different driving signals (i.e. the slope of the underlying PSD of the light curve is randomly varied to be between 1 and 3, and the seed for generating $a(t)$ is also randomly varied).

\subsection{The REV+PROP+NG model}

Now, we investigate the effect of random noise by introducing uncorrelated variability to the light curves. The Gaussian noise is added to the light curve model under different conditions of SNR, following the code provided by \cite{SNR_generator}. The light curve equation can be written by
\begin{equation}
 s(t) = a(t) + R \; a(t) \otimes \psi(t) + P \; a(t) \otimes {\rm TH}(t) + \alpha \; n(t) \;.
 \label{eq:s-rev+prop+n}
\end{equation}
According to Equations~\ref{eq:s/n} and~\ref{eq:s-rev+prop+n}, $\alpha$ and SNR can be fixed or free to serve different purposes. We fix $\alpha$ to determine the SNR in Equation~\ref{eq:s/n}, while in Equation~\ref{eq:s-rev+prop+n} $\alpha$ is instead determined as a function of SNR. We produce the light curves with various SNR scenarios, ranging from 0.1 to 1000, by varying $\alpha$ as a mixture weight to obtain the certain designated value of SNR.

\subsection{The REV+PROP+NG+NP model}

Further additional feature to the light curve is Poisson noise. The motivation is to mimic the readout process of CCD to make the simulated light curves more realistic. Light curves are resampled using built-in \texttt{NumPy} \texttt{random.poisson} function. These light curves are treated as the new signal where the Gaussian noise is also included with the similar method as mentioned in the previous section.

The light curves from the REV+PROP+NG+NP model are used to train the model that is applied to IRAS~13224--3809 data, as they closely resemble the observed ones. Example of light curves with the SNR correspond to the observed data is shown in Appendix~\ref{app:SNR}. All other setups for the modeled light curve are consistent with those in the REV+PROP+NG model, except that we extend the reflection fraction range to $R \in [1,10]$. Note that, in this study, we adopt simplified assumptions by fixing certain parameters (e.g., the photon index, black hole spin, inclination, and disk density) and employing a lamp-post source with a flexible free parameter, $R$. This approach generates disk response functions with different dilution effects (different $R$), possibly induced by variations in other parameters that are held constant. While a true composite spectral and timing model may be difficult to resolve, these choices help avoid an excessive number of potentially degenerate parameters. 

Nevertheless, we also show in Appendix~\ref{app:perfeval} the case when the model is trained with varying $R$, while fixing, e.g., $\Gamma$, $i$, and $M$, can still predict $h$ of the lamp-post source when testing with the new data generated from a ‘strict’ lamp-post case where we do not employ the variable $R$ but instead generate the data by varying $\Gamma$, $i$, and $M$ within the uncertainty of previously observed values for IRAS 13224-3809. This suggests that the current $R$ range is sufficient when applying the model to the specific AGN IRAS~13224--3809, as the test dataset is tailored to variations in $\Gamma$, $i$, $M$ within the observational uncertainties of this source.

\section{ML models for extracting the response functions}

\begin{figure}[h!]
        \centering
        \includegraphics[width = 0.8 \textwidth]{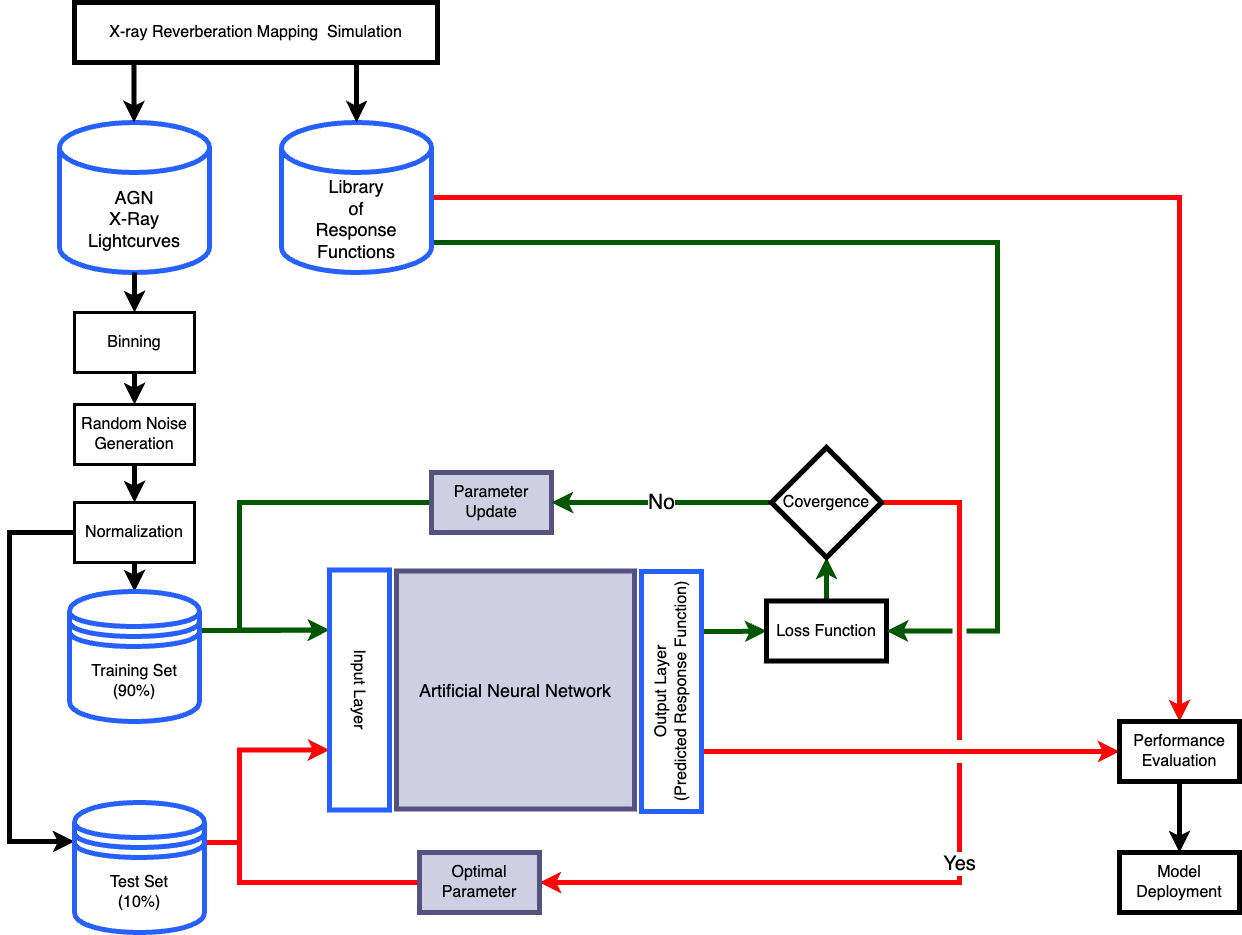}
        \caption{Generalized workflow for supervised ML. The light curves are binned with the bin size of 1~s, and 5~s for REV and REV+PROP models respectively. The green and red lines represent the pipe line for the optimization process and performance evaluation, respectively. See text for more details.}
        \label{fig:ANN_workflow}
    \end{figure}

The ML extraction pipeline involves supervised machine learning techniques, which is summarized in Figure~\ref{fig:ANN_workflow}. The data, which in our case are the simulated AGN light curves with X-ray reverberation features, are binned using 1~s and 5~s time bin for REV and REV+PROP models, respectively, and are separated into training and test sets by the fraction of 90\% of the training set and 10\% of the test set. Note that there is no missing data in the simulated light curves used to train the machine. Therefore, when applied to IRAS~13224--3809 data, we employ the Piecewise Cubic Hermite Interpolating Polynomial (PCHIP, \cite{PCHIP}) to replace the missing data and read-out error with estimated values. Training data are fed to the model, as it will learn to find the optimal transformation from input data to designated output feature (green line), by achieving the global minimum of the loss function via the numerical optimization process. Once the global optimum is reached, performance evaluation will take place by implementing the model to the test set to investigate whether the model is overfitted (red line). If it does not, the model is ready for deployment on the real observational data.

\subsection{Autoencoder (AE)}

The mathematical aspect of the AE is based on the multilayer perceptron (MLP) that works by mapping the input feature (\textbf{x}: light curve, in case of our work) to the designated output feature space \citep{Goodfellow-et-al-2016}, while each light curve consists \texttt{INPUT\_DIM} data points, which refers to the length of simulated light curve (\texttt{INPUT\_DIM} = 112400/\texttt{bin\_size} in our study). The \texttt{bin\_size} is assigned as 1~s for the REV model, and 5~s for all other models. A combination of simple functions provides the learning pipeline for the representation of input features in terms of the hidden layer (\textbf{h}). It can be simplified to the composition of linear mapping and activation functions, called fully connected layers, that can be mathematically represented by
    \begin{equation}
        \begin{split}
        \mathbf{h}_{1} &= f(\mathbf{A}_0\mathbf{x}+\mathbf{b}_0)\\
        \mathbf{h}_{2} &= f(\mathbf{A}_1\mathbf{h}_1+\mathbf{b}_1)\\
            &  \vdots\\
        \mathbf{h}_{n} &= f(\mathbf{A}_{n-1}\mathbf{h}_{n-1}+\mathbf{b}_{n-1})\\
        \mathbf{y}   &= f(\mathbf{A}_{n}\mathbf{h}_{n}+\mathbf{b}_{n}) \; .
    \end{split}   
    \label{eqn:mlp}
    \end{equation}
    
\begin{figure}[h!]
        \centering
        \includegraphics[width = 0.65 \textwidth]{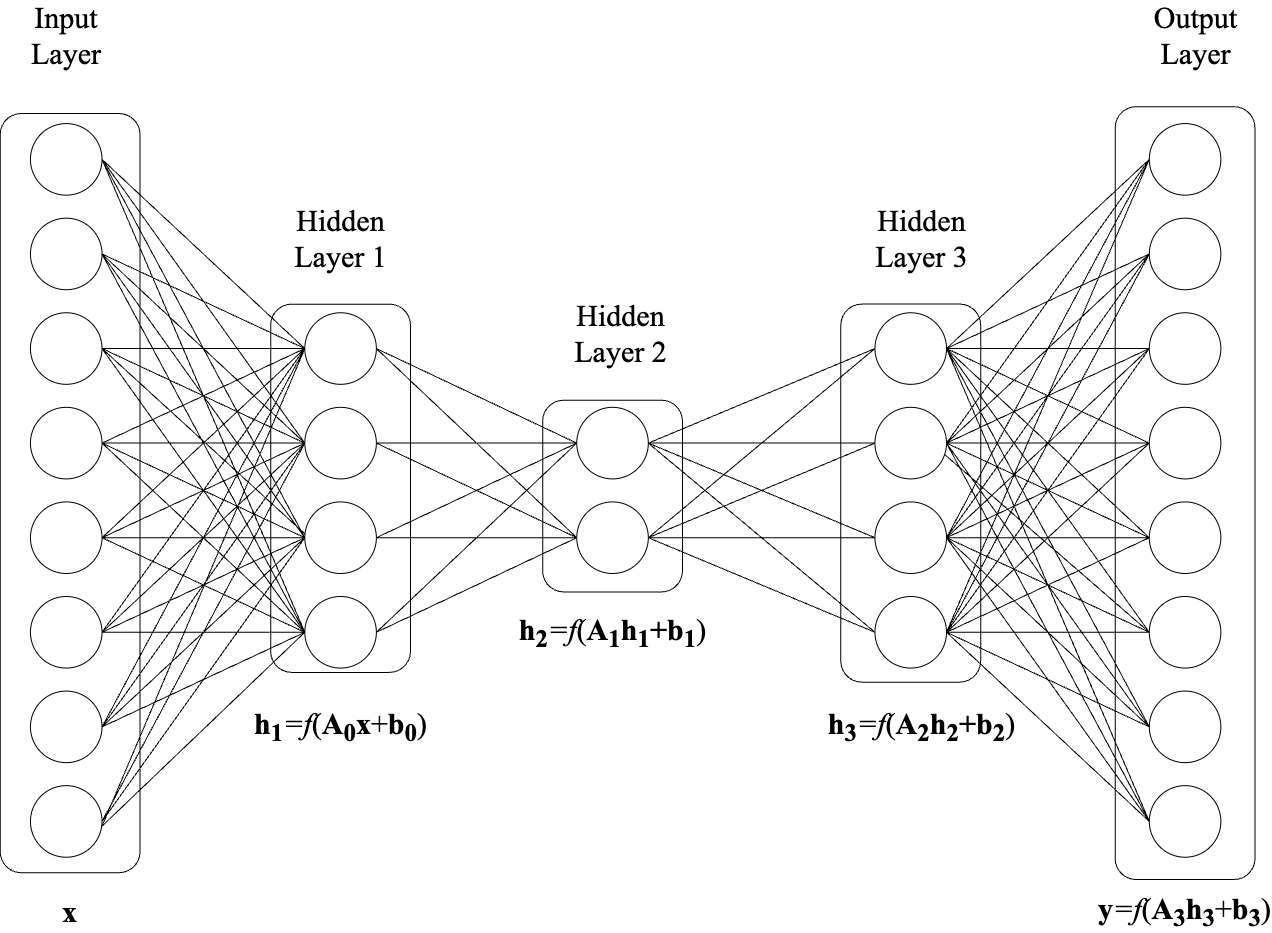}
        \caption{Example for autoencoding architecture with 3 hidden (fully-connected) layers. The network consists input layer, hidden layers, and output layer, resemble to Equation~\ref{eqn:mlp}}
        \label{fig:ae_network}
    \end{figure}

The network in Equation~\ref{eqn:mlp} can be illustrated in Figure~\ref{fig:ae_network}. Variables $\mathbf{A}$ and $\mathbf{b}$ are, respectively, the weight metric and the bias tensor, while their dimensions are determined by the number of input and output neurons for each layer. The function $f$ is the so-called activation function, which adds nonlinearity to the model and allowing the network to be suited to the complex system. It is often used in the form of hyperbolic tangent ($\tanh$), rectified linear unit (ReLU), or sigmoid function. Finally, target layer (\textbf{y}: response function); with length \texttt{OUTPUT\_DIM} = 10000 in our study, shall be returned from the output layer.

We implemented shallow autoencoder via PyTorch \citep{paszke2019pytorch}, consisting of 2 concatenated MLP networks, encoding layers, and decoding layers. Its architecture is shown in Table~\ref{tab:MLP}. The optimization process is performed to obtain the best-fit parameters for the network, subjected to the reconstruction loss. For example, the loss functions widely used in this architecture are L1 loss (Euclidean distance), mean square loss and cross-entropy loss. The optimization rate is restricted by the user-defined learning rate, to avoid overfitting and guarantee the convergence to the local minimum.

    \begin{table}
        \centering
        \caption{Architecture for standard AE. The user must provide the number of data points for each input feature (\texttt{INPUT\_DIM}), as well as for the output feature (\texttt{OUTPUT\_DIM}).}
        \begin{tabular}{|l|c|}
        \hline
             \textbf{Layer} &  \textbf{ Input $\times$ Output}   \\
             \hline
             Input Layer    &   \texttt{INPUT\_DIM} $\times$ 400 \\
             Hidden Layer 1 &   400 $\times$ 400 \\
             Hidden Layer 2 &   400 $\times$ 200 \\
             Hidden Layer 3 &   200 $\times$ 400 \\
             Hidden Layer 4 &   400 $\times$ 400 \\
             Output Layer   &   400 $\times$ \texttt{OUTPUT\_DIM}\\
        \hline
        \end{tabular}
        \label{tab:MLP}
    \end{table}

\subsection{Variational Autoencoder (VAE)}
Mathematical expression of VAE can be written in a similar way as Equation~\ref{eqn:mlp}, comprise two consecutive MLP networks as encoding and decoding layers. However, VAE is interposed by the reparameterization process. The encoder transforms the input feature into a lower dimension, achieving a mutual representation of the transformed features of the arithmetic mean ($\mu$), and variance ($\sigma^2$), as the latent space. The decoder then transforms the resampled latent space by increasing the dimension of the feature to the designated output shape. Reparameterization is performed to resample the latent space, resulting in the latent vector $\mathbf{z}$, which is recovered from encoded $\mathbf{\mu}$, $\mathbf{\sigma}$, and additional random number ($\epsilon$), derived from a Gaussian probability distribution, with respect to Equation~\ref{eqn:vae},
    \begin{equation}
            \mathbf{z} = \mathbf{\mu} + \epsilon\;\mathbf{\sigma} \;.
        \label{eqn:vae}
    \end{equation}

We also use the PyTorch library and adopt the code from \cite{Kang_2021} to construct the VAE as Table~\ref{tab:VAE}. The optimization process is governed by a linear combination of binary cross-entropy loss ($BCE$), to determine how well the response functions are predicted ($\hat{y}(t)$) compared with its ground truths ($y(t)$), and Kullback-Leibler divergence loss ($KL$), to penalize the performance of reparameterization process by considering the behaviour of $\mathbf{\mu}$ and $\mathbf{\sigma}$ in latent space, with time-wise weight for each point of response: 
     
     \begin{equation}
         \begin{split}
             BCE(\hat{y},y) &= -\sum_i w_i \Big( y_i \log(\hat{y}_i) + (1-y_i) \cdot \log(1-\hat{y}_i) \Big) \;,    \\
             \text{s.t.} &\sum_i w_i = 1 \;,
         \end{split}
         \label{eqn:bce_loss}
    \end{equation}
     \begin{equation}
     KL(\mu,\sigma) = -\frac{1}{2} \sum_j \Big( 1 + \log(\sigma_j) - \mu_j^2 - \sigma_j \Big)   \;.
     \label{eqn:kl_loss}
    \end{equation}
    An index $i$ is the time-stamp index that runs from 1 to \texttt{OUTPUT\_DIM}, while $j$ is indexing every element in representation tensors $\mathbf{\mu}$ and $\mathbf{\sigma}$ ($j$ runs between 1 to 200), as defined in Table~\ref{tab:VAE}. Combining Equations~\ref{eqn:bce_loss} and~\ref{eqn:kl_loss}, the loss function for VAE can then be written as
     \begin{equation}
     \mathcal{L} = BCE(\hat{y},y) + KL(\mu,\sigma) \;.
    \end{equation}

    \begin{table}
        \centering
        \caption{VAE composition. The network performs a combination of linear mapping to translate the input feature to the target vector. Note that the VAE assumes the Gaussianity behavior in the data. Per mathematical definition, features computed by encoding layers shall be represented in terms of mean ($\mathbf{\mu}$) from Mean layer, and variance ($\mathbf{\sigma^2}$) from Variance layer, respectively. Mean and Variance layer encoded the yield of hidden layer from length 400 to 200. See text for more details.}
        \begin{tabular}{|l|c|}
        \multicolumn{2}{c}{\textbf{Encoding Layers}}\\
        \hline
             \textbf{Layer} &  \textbf{ Input $\times$ Output}   \\
             \hline
             Input Layer    &   \texttt{INPUT\_DIM} $\times$ 400 \\
             Hidden Layer   &   400 $\times$ 400 \\
             Mean           &   400 $\times$ 200 \\
             Variance       &   400 $\times$ 200 \\
        \hline
        \multicolumn{2}{c}{}\\
        \multicolumn{2}{c}{\textbf{Decoding Layers}}\\
        \hline
            \textbf{Layer} &  \textbf{ Input $\times$ Output}   \\
            \hline
            Hidden Layer 2 & 200 $\times$ 400 \\
            Hidden Layer 3 & 400 $\times$ 400 \\
            Output Layer   & 400 $\times$ \texttt{OUTPUT\_DIM}\\
        \hline
        \end{tabular}
        \label{tab:VAE}
    \end{table}

\subsection{Signals similarity indicator}
    
In order to verify the closeness between a true response function and one predicted by the ML model, statistical indicators are used to determine the proximity between the predicted curves ($\hat{y}(t)$) and their ground truths ($y(t)$). Standard metrics such as R$^2$-score, mean absolute error (MAE), and mean square error (MSE), can also be used to measure the performance of prediction. However, we instead use cross-correlation and signal coherence as the performance metrics, as they are widely used to estimate similarity between two time-series data. The cross-correlation (CC) is the measurement of the similarity between two time series under different time delays, with the governing equation:
    \begin{equation}
        (\hat{y} \star y)(\tau) = \int_{-\infty}^{\infty} \overline{\hat{y}(t)}y(t-\tau) dt \;,
    \end{equation}
where $\overline{\hat{y}(t)}$ denotes the complex conjugate of $\hat{y}(t)$. We use the \texttt{scipy.signal.correlate} function to determine the correlation between the signals. Only zero-lag ($\tau=0$) cross-correlation is considered to determine the prediction accuracy of the model.

On the other hand, the signal coherence is also used to estimate the relationship between two signals $\mathbf{\hat{y}}$ and $\mathbf{y}$ in a Fourier space, as the likelihood between two power spectra. The equation is defined as:
    \begin{equation}
        C_{\hat{y}y}(f) = \frac{\vert G_{\hat{y}y}(f) \vert ^2}{G_{\hat{y}\hat{y}}(f)G_{yy}(f)} \;.
    \end{equation}
$G_{\hat{y}y}$ is cross-spectral density between $\mathbf{\hat{y}}$ and $\mathbf{y}$ while $G_{\hat{y}\hat{y}}$ and $G_{yy}$ are the autospectral density of $\mathbf{\hat{y}}$ and $\mathbf{y}$, respectively. In this work, the built-in function \texttt{scipy.signal.coherence} is used to compute the coherence at zero Hertz component; as the response functions are comprised of low-frequency elements.

\begin{figure}[h!]
        \centering
        \begin{subfigure}{\textwidth}
            \centering
            \includegraphics[width = \linewidth]{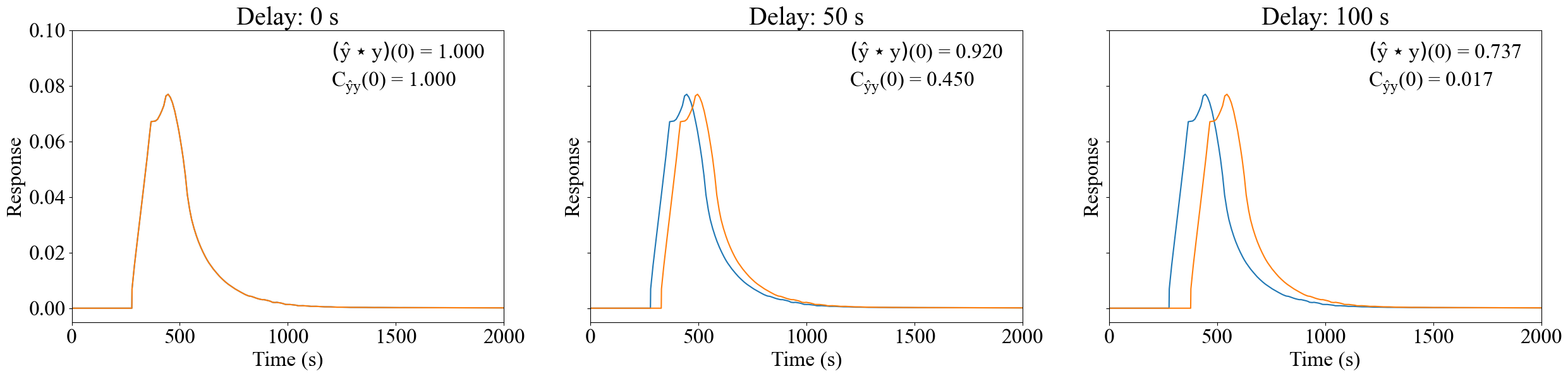}
            \label{subfig:signal_similarity_1}
        \end{subfigure}
        
        \begin{subfigure}{\textwidth}
            \centering
            \includegraphics[width = \linewidth]{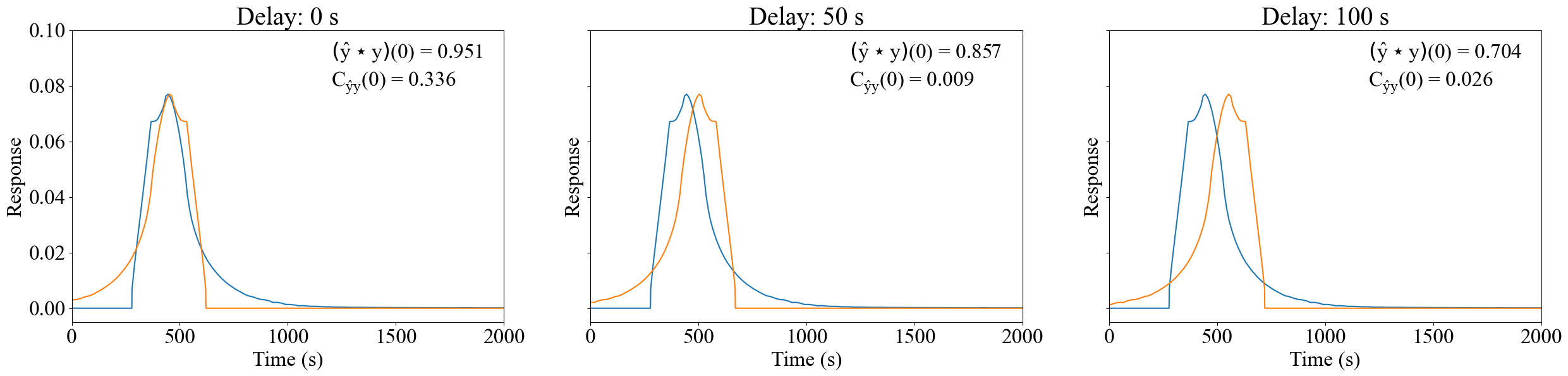}
            \label{subfig:signal_similarity_2}
        \end{subfigure}
        \caption{Similarity measurements (cross correlation and signal coherence) of mock response functions under different conditions. Top panels: two response signals are different only by a specific time shift. Bottom panels: two responses are different by the time shift and one is also the mirror image of the other. See text for more details.}
        \label{fig:enter-label}
    \end{figure}

Examples of the similarity measurements using the cross-correlation and signal coherence are presented in Figure~\ref{fig:enter-label}. We produce mock response functions different in shape and response time. It can be seen that the signal coherence seems to drop faster especially when their shapes are mismatched.

\section{Results}

    \subsection{Model performance}

     To evaluate the performance of VAE, we set up three simultaneous experiments that handle different sources of data, comparing with standard AE as a baseline. To control the computation, we define the network to use a similar amount of trainable parameters, together with Adam optimizer at a learning rate of $10^{-3}$ and a training process of 2000 epochs. 

\begin{figure}[h!]
        \centering
        \includegraphics[width = 0.95 \textwidth]{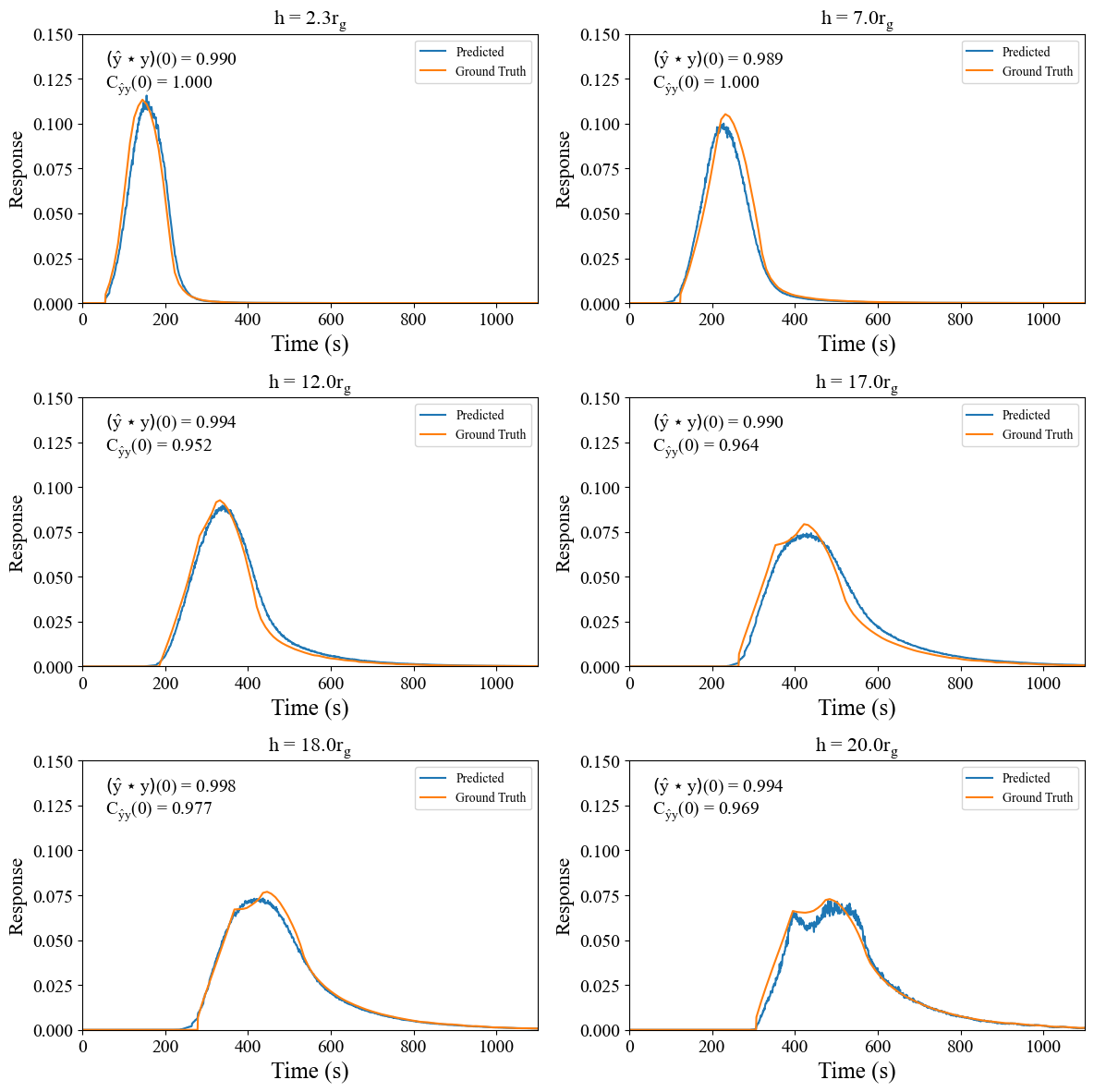}
        \caption{Example of the X-ray reverberation response functions predicted by AE (blue lines) compared with actual \textsc{kynxilrev} response functions (orange lines) for the REV+PROP model.}
        \label{fig:reconstruction}
    \end{figure}

     Table~\ref{tab:REV_full} shows the accuracy results in the cases of noise-free light curves generated by the REV and REV+PROP model, using driving signals with randomized geometrical parameters. It can be seen that, for the noise-free light curves, the accuracy of both AE and VAE models to extract the reverberation response function are quite high (CC and signal coherence is more than $\sim 0.9$) whether or not the effects due to the disc propagating fluctuations are included. The examples of the extracted response functions compared to the real response profiles for The REV+PROP model are also presented in Figure~\ref{fig:reconstruction}. The models perform well for a broad range of coronal heights. The shape of the response function is best recovered in the case of $h = 20~r_{\rm g}$, as can be seen from the distinct double-peak feature which consists of a small peak at $\sim 400$~s and a broader peak at $\sim 500$~s. These double-peak features are not clearly visible in the predicted response functions for $h=$18, 17, and $12~r_{\rm g}$. For $h=2.3$ and $7~r_{\rm g}$, the recovered response functions are good with slightly shifted to larger delays and to smaller delays, respectively.

    \begin{table}[H]
        \centering
        \begin{tabular}{|c|c|c|c|c|c|c|c|c|}
            \hline
                \multirow{3}{*}{Architecture} & \multicolumn{4}{c|}{REV model} & \multicolumn{4}{c|}{REV+PROP Model}\\
            \cline{2-9}
                & \multicolumn{2}{c|}{Cross-Correlation} & \multicolumn{2}{c|}{Signal Coherence} & \multicolumn{2}{c|}{Cross-Correlation} & \multicolumn{2}{c|}{Signal Coherence}\\
               \cline{2-9}
                & Training Set & Test Set & Training Set & Test Set & Training Set & Test Set & Training Set & Test Set\\
             \hline
            
             AE          & $1.000^{+0.000}_{-0.000}$ & $0.987^{+0.010}_{-0.055}$ 
             & $1.000^{+0.000}_{-0.000}$ & $0.971^{+0.028}_{-0.287}$
             & $1.000^{+0.000}_{-0.000}$ & $0.979^{+0.016}_{-0.077}$ & $1.000^{+0.000}_{-0.003}$ & $0.953^{+0.041}_{-0.267}$
             \\ 

             VAE         & $0.992^{+0.006}_{-0.014}$ & $0.983^{+0.013}_{-0.046}$ 
             & $0.988^{+0.011}_{-0.056}$ & $0.959^{+0.037}_{-0.195}$
             & $0.990^{+0.007}_{-0.019}$ & $0.962^{+0.030}_{-0.081}$ & $0.980^{+0.018}_{-0.090}$ & $0.905^{+0.083}_{-0.339}$
             
             \\
             \hline
        \end{tabular}
        \caption{Prediction likelihood for the response function extracted by training the autoencoders with simulated REV and REV+PROP model light curves.}
        \label{tab:REV_full}
    \end{table}

\begin{figure}
        \centering
        \begin{subfigure}{ 0.47 \textwidth}
            \includegraphics[width = \linewidth]{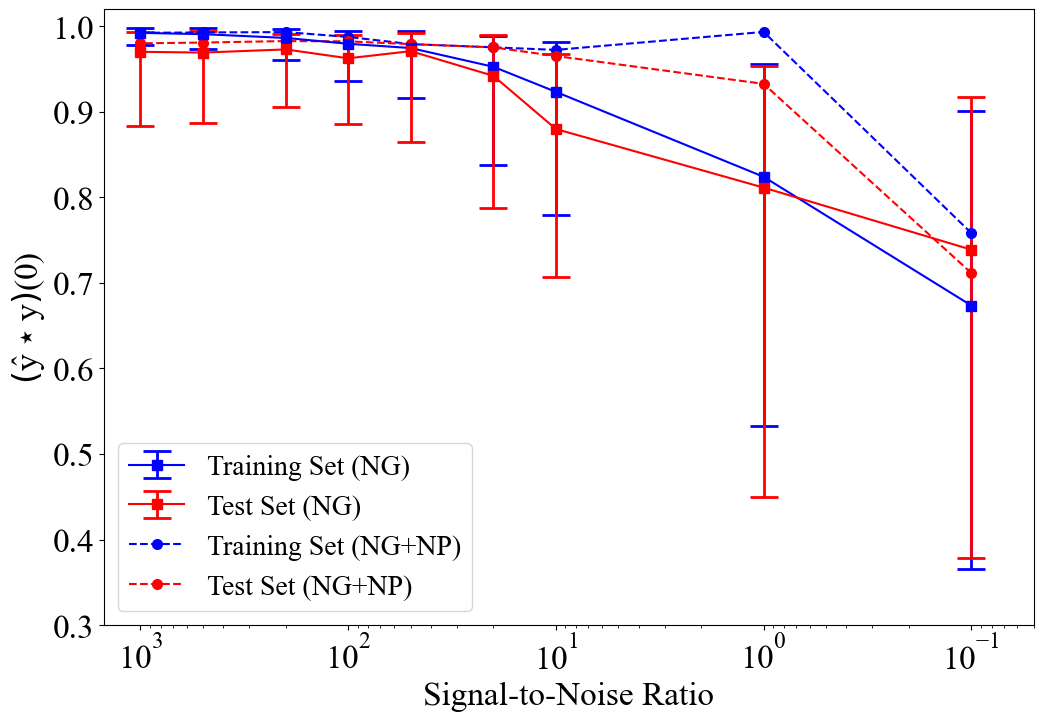}
        \end{subfigure}
        \begin{subfigure}{ 0.47 \textwidth}
            \includegraphics[width = \linewidth]{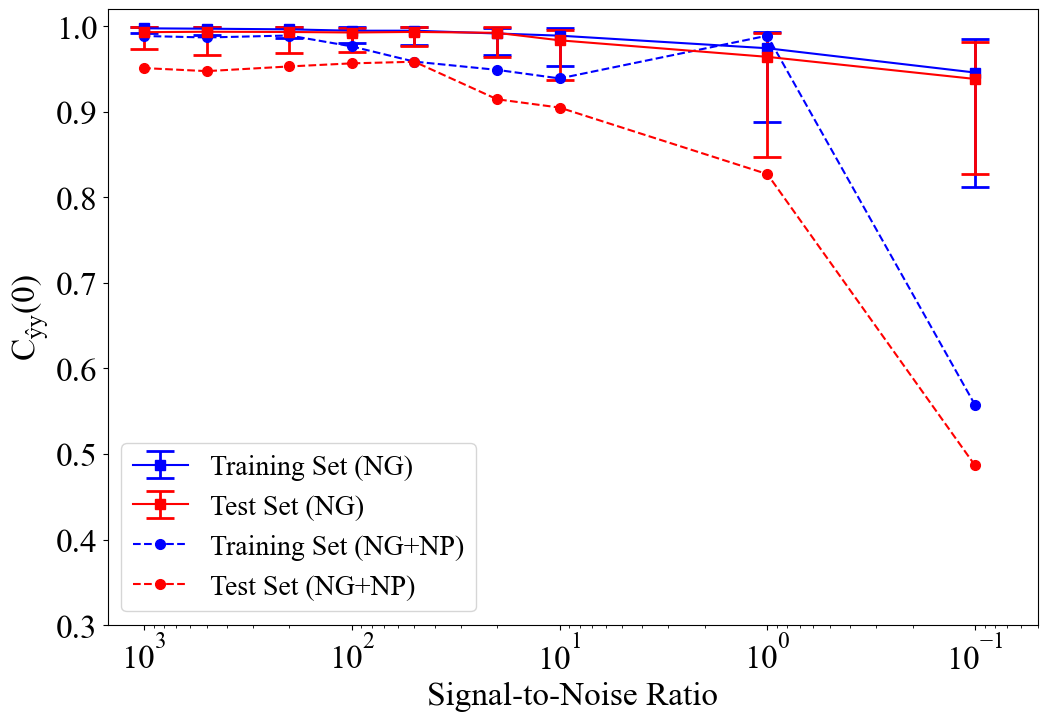}
        \end{subfigure}
        
        \caption{The evolution of response functions likelihood with the SNR, using the CC (left panel) and signal coherence (right panel) as similarity indicators. Solid lines with square markers are likelihood for the REV+PROP+NG model, while dashed lines with dotted markers are for the REV+PROP+NG+NP model, while red and blue color represent training and test set respectively. 
        When SNR decreases, the CC seems to drop more rapidly than the coherence. The error bars correspond to the interquartile range between the first and third quartile, in order to interpret how well the overall response functions are extracted.}
        \label{fig:sc_noise}
    \end{figure}

Now, we investigate the REV+PROP+NG model which is the case when random noises contaminated the light curve. The training and testing are performed under a signal-to-noise ratio ranging from 0.1 to 1000. Note that the uncorrelated noise is added after the binning process. How the accuracy decreases with increasing SNR is presented in Figure~\ref{fig:sc_noise}. We can see that both CC and signal coherence gradually decline when noise becomes more dominant. The trend of decreasing accuracy with decreasing SNR is quite comparable between the training and test datasets, meaning that the models are not overfitting. However, the accuracy measured by CC seems to drop more rapidly than the coherence. These results suggested that the models perform well (accuracy from both indicators is $\gtrsim 0.9$) if the SNR $\gtrsim 50$.

We also used VAE with Poissonian resampled light curves and additional Gaussian noise (REV+PROP+NG+NP model) in the identical SNR range to train and test model. Cross Correlation for training and test light curves agree at the accuracy of $\gtrsim 0.9$ at SNR $\gtrsim 10$. However, signal coherence gradually declines when SNR $\lesssim 50$. Thus, it could be interpreted that the prediction sensitivity of autoencoder for the REV+PROP+NG+NP model also agrees at accuracy of $\gtrsim 0.9$ and SNR $\gtrsim 50$, comparable to that of the REV+PROP+NG model.

We also evaluate the robustness of the model for the general applicability when the observed or test data have slightly different characteristics to the training data (i.e. have different $M$, $i$, and $\Gamma$). Results and discussion are presented in Appendix~\ref{app:perfeval}.

    \subsection{Model deployment on real data}

    In this part, we used 13 light curve observations from IRAS 13224-3809, with SNR greater than 50, to predict their respective response functions using the REV+PROP+NG+NP model. However, we emphasise here that the data SNR presented in this work was obtained from the definition in Equation~\ref{eq:s/n}, in which this is different from, and cannot be compared directly to,  the value obtained from the typical calculation (cf. table 1 of \citealt{Nakhonthong2024}). Indeed, while the value from the typical calculation is dependent on the numbers of source and background photons, the SNR determined in this work is, in principle, proportional to the total power of variability of the data, compared to the total power of frequency-independent, random (i.e., white) noise, regardless of the origin of variability whether it is from source or background. However, as the light curves used here are background-subtracted, we argue that the detected variability is dominated by the source's photons. In fact, determining the SNR in this way is appropriate in this context as this could ensure that the datasets used have a sufficient level of variability for our analysis. 
    
\begin{figure}[h!]
        \centering
        \includegraphics[width=0.95 \textwidth]{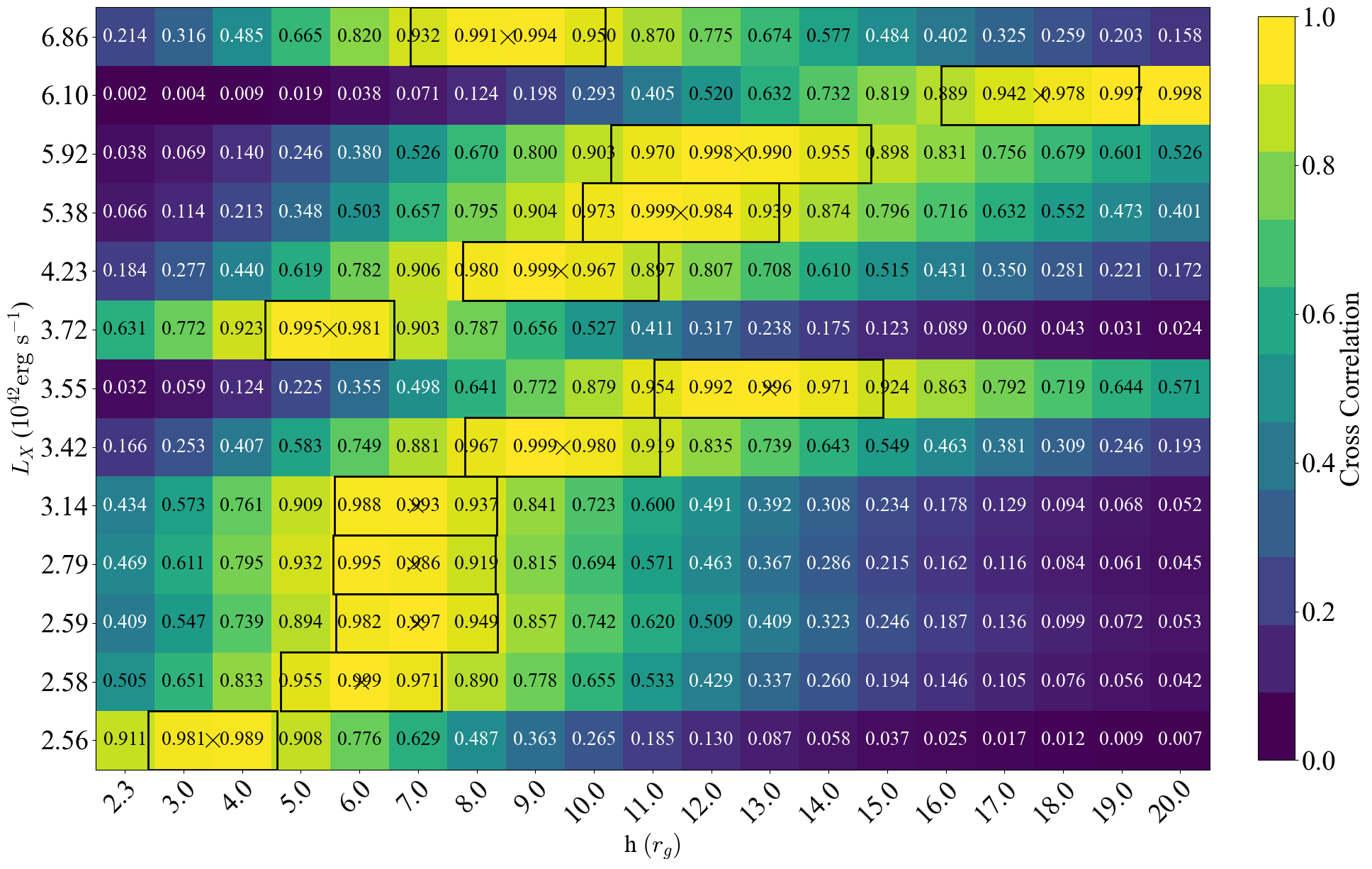}
        \caption{Matching score, by zero-lag cross-correlation, of the predicted response function using the REV+PROP+NG+NP model with the response function generated by \textsc{kynxilrev}. The colors used in this plot represent the scoring level with the lowest value of 0 (dark blue), and the highest value of 1 (yellow). Cross marks and boxes in the illustration imply, respectively, the expected coronal height and standard deviation for each response function computed by the eighty-percent rule.}
        \label{fig:real_data_deploy}
    \end{figure}

\begin{figure}[h!]
        \centering
        \begin{subfigure}[h]{0.50\textwidth}
            \centering
            \includegraphics[width = 1. \textwidth]{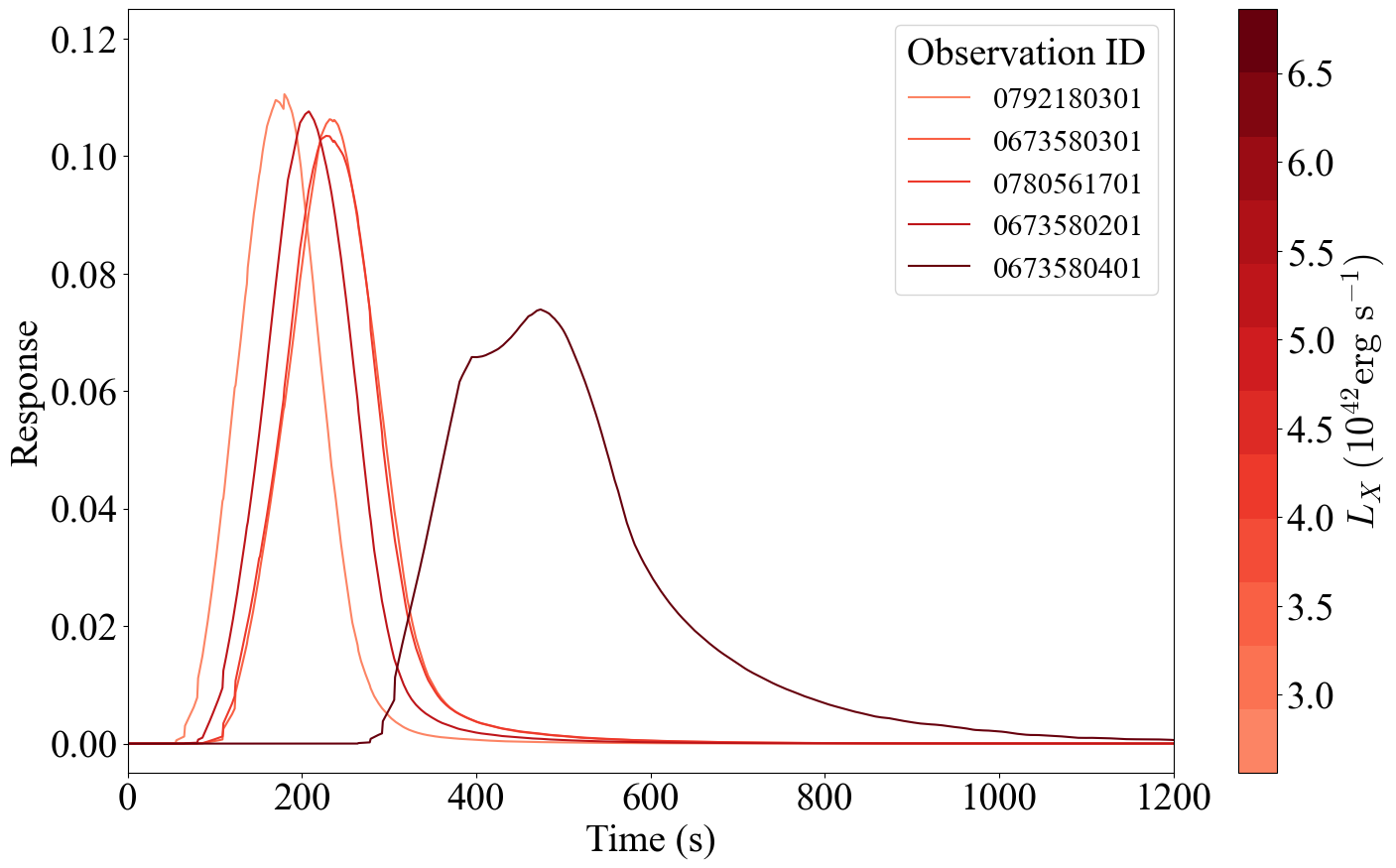}
        \end{subfigure}
        ~
        \begin{subfigure}[h]{0.45 \textwidth}
            \centering
            \includegraphics[width = 1.\textwidth]{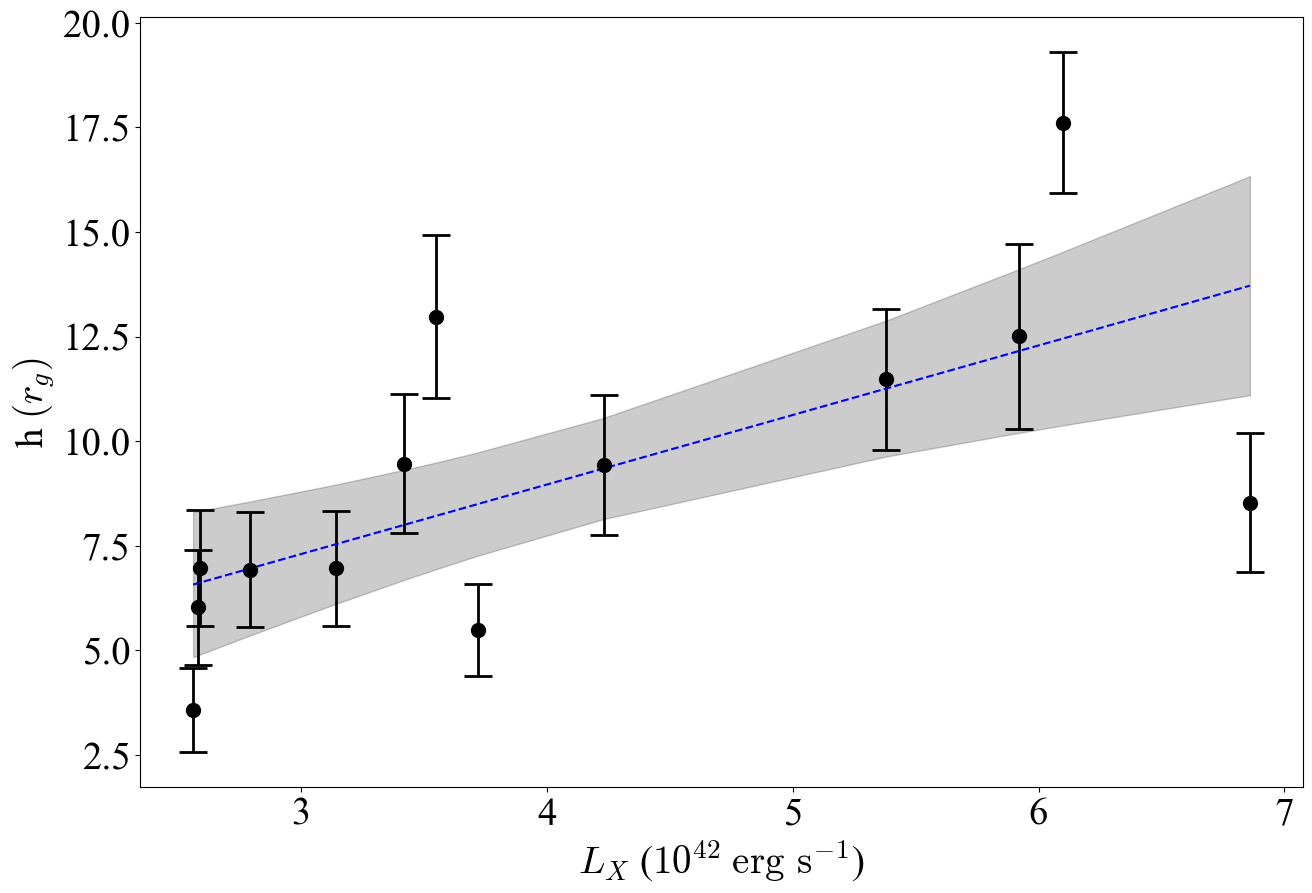}
        \end{subfigure}
        \caption{Left: Examples of the reconstructed response functions of some IRAS~13224--3809 observations using the REV+PROP+NG+NP model. The colors of the profiles show the source luminosity measured in 2--10~keV band ($L_{X}$). Right: Predicted coronal heights ($h$) plotted against $L_{X}$ in 13 observations of IRAS~13224--3809. Blue dotted line indicates the linear regression prediction of the coronal height with respective X-ray luminosity, while 3$\sigma$ confident interval is illustrated by grey region.}
        \label{fig:real_data_deloy_prediction}
    \end{figure}

    Note that, prior to prediction, we use Piecewise Cubic Hermite Interpolting Polynomials (PCHIP) to replace the missing values or gaps in the IRAS~13224--3809 data with an estimate. We compare the predicted result with the category of response functions of a lamp-post corona situated at the height between 2.3~$r_{\rm g}$ and 20.0~$r_{\rm g}$. We investigated the likelihood between each predicted response and prescribed response using cross-correlation similarity. Thereafter, the expected scale height and its standard deviation shall be computed with eighty-percent rule adopted from \cite{nunez_gaussian_2023}, as shown in Figure~\ref{fig:real_data_deploy}. In other words, only coronal heights; yielding cross-correlation $\gtrsim 0.8$ of maximum cross-correlation to each predicted response, are considered. The factor of 0.8 has been conventionally used in time delay measurement, as mentioned in \cite{Peterson_2004}. This criterion is adopted to this work to bestow the prediction uncertainty of the coronal height. Note that thresholding at 0.9 or 0.95 would leave only a few values in computation leading to small uncertainty. On the other hand, a smaller threshold at, e.g., 0.7 would lead to larger uncertainty, with a slight change of expected coronal scale height. The monotonic relationship between X-ray luminosity and coronal height obtained here is comparable with, e.g., \cite{Alston2020}, as shown in Figure~\ref{fig:real_data_deloy_prediction}, with the Spearman correlation coefficient of 0.68 (p-value: 0.011).

\begin{figure}[h!]
        \centering
        \begin{subfigure}[h]{0.48 \textwidth}
            \centering
            \includegraphics[width= 1.\textwidth]{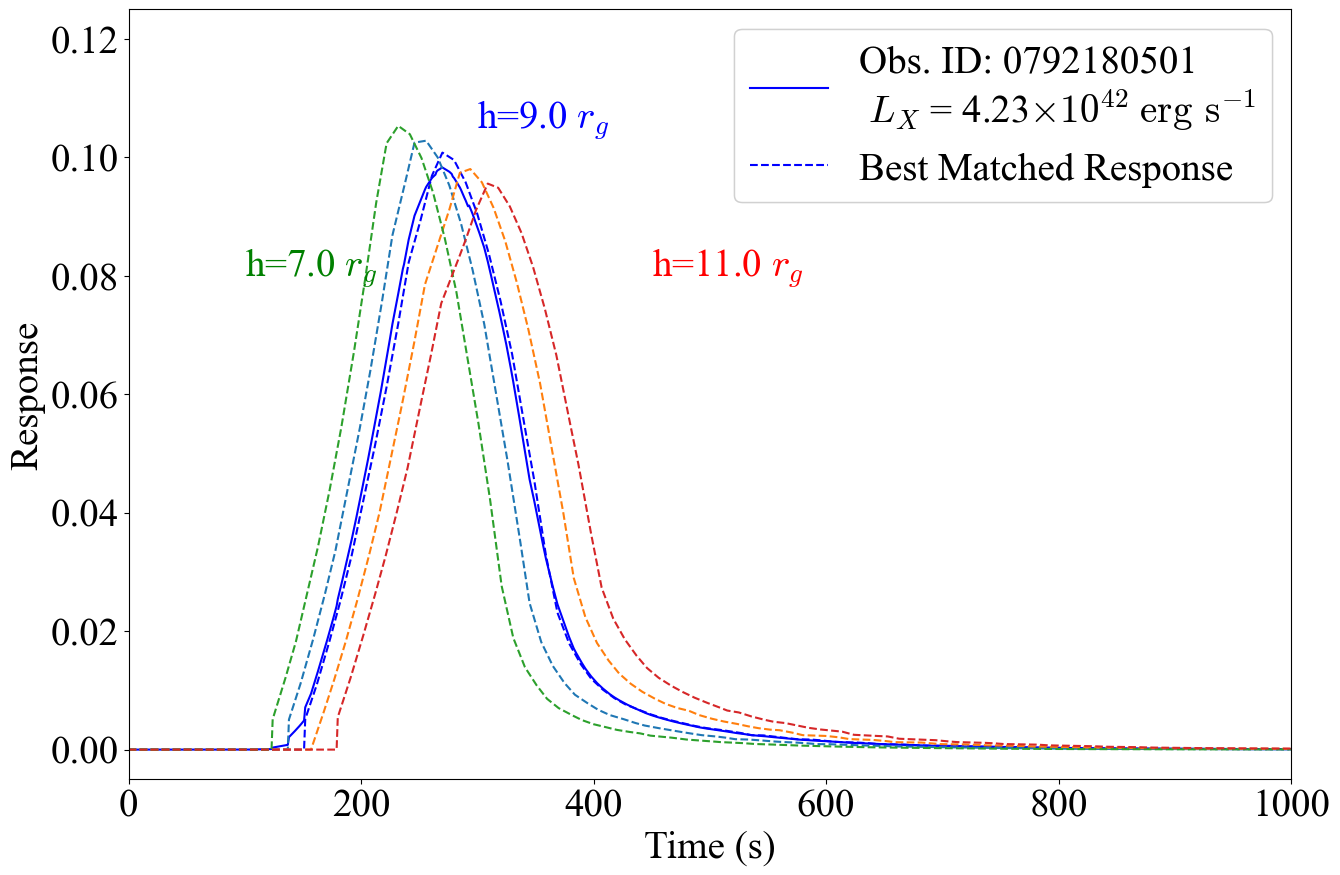}
        \end{subfigure}
        ~
        \begin{subfigure}[h]{0.45\textwidth}
            \centering
            \includegraphics[width=1.\textwidth]{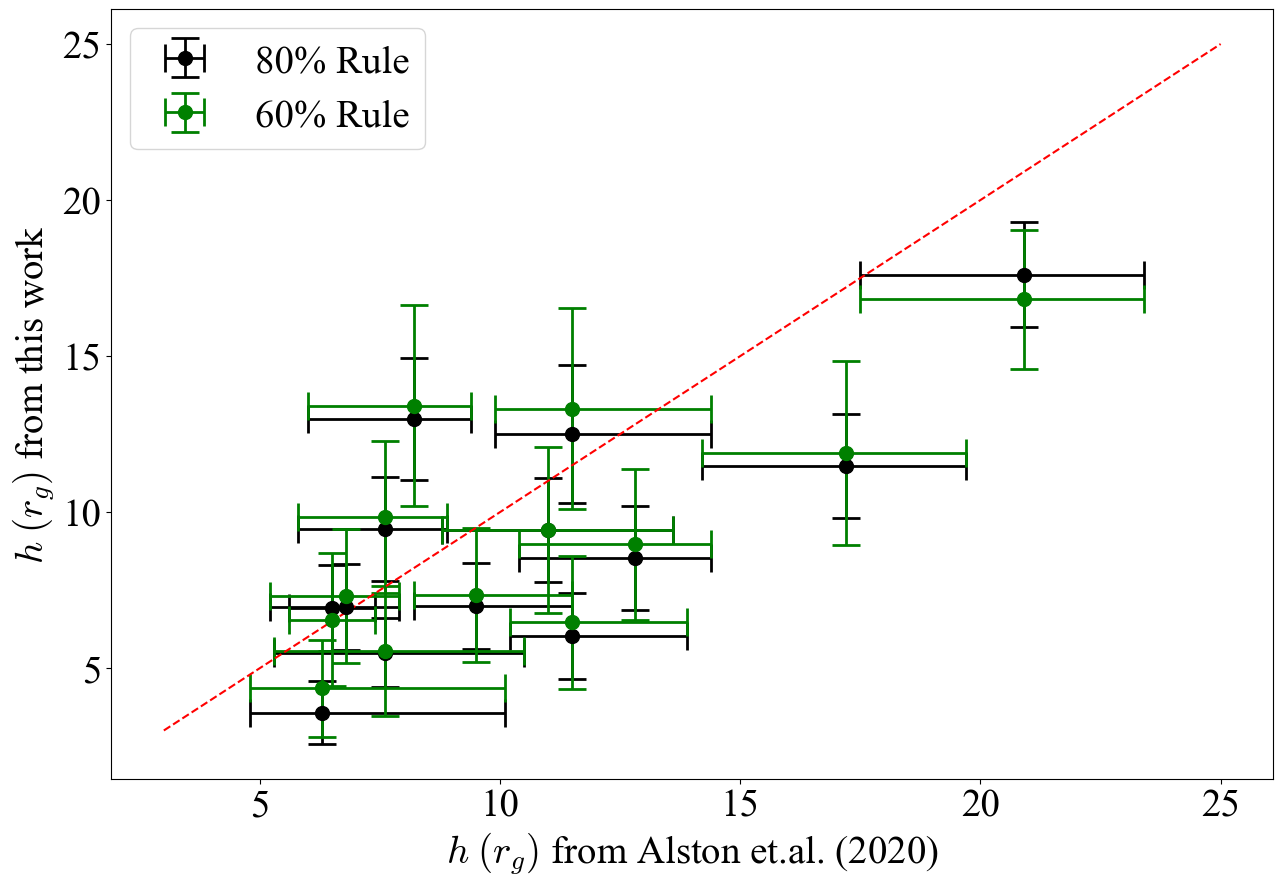}
        \end{subfigure}
        \caption{Left: Predicted response functions of IRAS~13224--3809 (Obs. ID 0792180501; blue-solid line) and some \textsc{kynxilrev} response functions corresponding to different coronal heights that also show cross-correlations $> 0.8$ (dotted lines). The best-matched response in this case is $h=9~r_{\rm g}$ (blue-dotted line). Right: Prediction of coronal height by our model compared with what recovered by \cite{Alston2020}. The 1--1 line is marked as red-dashed line. Prediction uncertainty by 80\% rule (black crosshair) and 60\% rule (green crosshair) are illustrated. We note that our approach uses only a single energy band to estimate $h$, with $\Gamma$ fixed at 2, which may limit the ability to fully capture the source's spectral evolution. Nevertheless, considering the significant uncertainties in the constrained $h$, our results remain broadly consistent with those of \cite{Alston2020}.}
        \label{fig:eighty_vs_sixty}
    \end{figure}

    Figure.~\ref{fig:eighty_vs_sixty} (left panel) shows, as an example, the predicted response function for IRAS~13224–3809 (Obs. ID 0792180501). We illustrated the predicted response in a blue-solid line, while its best-matched ground-truth response corresponding to the case of $h = 9~r_{\rm g}$ is illustrated in the blue-dotted lines. Other ground-truth responses that can lead to cross-correlations more than 0.8 are also presented in the dotted lines in different colors with the lowest and highest coronal height of $7~r_{\rm g}$ and and $11~r_{\rm g}$
    
    Figure.~\ref{fig:eighty_vs_sixty} (right panel) shows the comparison between the height estimates for each observation in this work to those obtained in \cite{Alston2020}. We also compare the cases when the standard deviation is computed using the threshold levels of 0.8 and 0.6. The constrained heights are loosely comparable within the estimated errors. While there is no clear trend compared to 1-1 line, the individual mean values show some discrepancies which might be because different assumptions are used in the models. For example, we fix $\Gamma$, while in \cite{Alston2020} it is derived from the 0.3--10 keV energy spectrum model fits and is varied among different observations. Here, our approach only employ the soft-band (0.3--1 keV) light curve to derive the coronal height.

\section{discussion and conclusion}

Analyses of X-ray variability in AGN have typically been limited to Fourier space (e.g., by measuring the power spectral density and the lag spectrum). The X-ray reverberation time delays are usually measured from the phases of the cross-spectrum \citep[e.g.][]{Fabian2009, Zoghbi2010, Wilkins2013, Cackett2014, Emmanoulopoulos2014, Chainakun2015, Epitropakis2016}, and no studies have been able to extract the associated response function directly from an AGN light curve. This work aims to tackle the problem in the time domain by developing the VAE model, which enables us to directly derive the response functions from the light curve. Subsequently, this technique offers an independent tool for mapping the disc-corona geometry straightforwardly from the estimated shape of the reverberation response function.

VAE is the neural network architecture sharing a similar mathematical basis to the multilayer perceptron~\citep{Goodfellow-et-al-2016, paszke2019pytorch,Kang_2021}, which is supervised learning. The network learns the transformation between light curves and response functions via numerical optimization. Using the standard autoencoder as the baseline, we find that VAE performed well in predicting the response functions using noise-free light curves generated by REV and REV+PROP models. We also examine the limitation of VAE by adding uncorrelated Gaussian noise to generated light curves. The SNR value is determined by the power spectral fraction between the signal and additional noise. The results show that the prediction accuracy declines gradually when SNR $< 50$. Therefore, we deploy VAE on IRAS 13224--3809 light curves with SNR $\gtrsim 50$. Data is preprocessed by binning and then interpolation to approximate the photon count in the gap area, as well as normalization, before the prediction of response functions. The best-matched response functions are determined by comparing the predicted response with \textsc{kynxilrev} simulated response functions in the archive. Only candidates with zero-lag cross-correlation $\gtrsim$ 0.8 are used to compute the expected coronal height and its standard deviation. The criteria are adopted from the eighty-percent rule mentioned in~\cite{nunez_gaussian_2023}. 

With the developed VAE model, we can obtain an estimate of the shape of the reverberation response function of IRAS~13224--3809. Our findings demonstrate a dynamic corona changing its height with luminosity, from $\sim 3$--20~$r_{\rm g}$, which aligns with previous studies. \cite{Alston2020} discovered, through an analysis of the lag-frequency spectra, that the lamp-post coronal height increases from $\sim 6$--20~$r_{\rm g}$ with source luminosity. By assuming $a=0.99$ and analyzing combined spectral-timing data in various flux states, \cite{Caballero2020} found a trend for the source height rising from $\sim 3 \ r_{\rm g}$ to $10^{+10}_{-1} \ r_{\rm g}$ with luminosity. Furthermore, due to the gravitational light-bending effects, the reflection component in IRAS~13224--3809 is less variable compared to the primary X-ray continuum \citep{Miniutti2004, Chiang2015}. The reverberation effects then act as a filter that suppresses the fractional excess variance of intrinsic variability, producing the reverberation dip on the PSD profiles whose amplitude is stronger in the energy band that is more dominated by reflection \citep[e.g.][]{Emmanoulopoulos2016, Papadakis2016, Chainakun2019a}. The coronal height correlated with luminosity was also suggested by the studies of these reverberation dips on the PSD data of IRAS~13224--3809 \citep{Chainakun2022b, Mankatwit2023}. This work then provides further evidence supporting such scenario.

While we use {\sc kynxilrev}, a part of the {\sc kyn} package \citep{Dovciak2004a, Dovciak2004b, Caballero2018}, that employs the {\sc xillver} reflection model \citep{Garcia2010,Garcia2013} to generate the disk response functions, it is also possible to use {\sc kynrefrev} that relies on the {\sc reflionx} reflection model \citep{Ross1999, Ross2005}. Different choices of these models may lead to different dilution effects on the amplitude of the lags, resulting in a bias in determining the coronal height \citep{Garcia2013, Khanthasombat2024}. The alternative model for the reflection from a high-density disk is also possible for IRAS~13224--3809 \citep{Jiang2022}. We predict the response function using only soft X-ray light curves for both training and deployment, whereas \cite{Alston2020} incorporates both soft and hard X-rays for time-lag analysis. Additionally, there is evidence that $\Gamma$ might be variable and flux-dependent in IRAS~13224--3809 \citep[e.g.][]{Alston2020, Caballero2020}. To maintain simplicity, we initially set $\Gamma=2$ and employ free parameter R to compensate for the dilution effects possibly induced by variations in those parameters that are held constant. Although this simplified analysis process may overlook certain complex spectral behaviors, the $h$--$L_{\rm X}$ correlation trend can be seen, and our results are not drastically different from what obtained by, for example, \cite{Alston2020}. The uncertainties in coronal height measurements across different methods emphasize the need for more advanced measurement techniques, cross-validation between methods, and higher-quality observational data to better constrain the models. At the moment, the model tries to fit the overall shape of the response rather than locally concentrating on the feature surrounding the peak. Nevertheless, judging by the overall shape of the reconstructed response functions, $h$ should still be accurately predicted by the model as long as SNR~$\gtrsim 50$. Predicting various parameters at once may require the model to fully capture the complex details around the response peak, which is subject to future study.

Nevertheless, we expect the model to work well even if it is designed to predict the TH-shaped response functions instead. In this case, it could be used to test the Comptonisation models or disk-propagation models. Each of these delays should be characterized by an energy-dependent response function. \cite{Jaiswal2023} investigated the time delays in AGNs caused by both accretion disk and broad-line region (BLR) reprocessing. They concluded that the effects of BLR scattering behave similarly to those resulting from an increase in the height of the X-ray source above the disk. This introduces degeneracy that complicates the interpretation of time lags and may also lead to overestimated coronal heights. However, \cite{Nunez2023} have shown that the primary source of contamination contributing to the increased accretion disk time delay is predominantly related to diffuse continuum emission from the BLR \citep{Chelouche2019}, with little or negligible dependence on the height of the X-ray source. Therefore, the presence of this degeneracy remains a matter of debate. The lags in IRAS~13224--3809 may be explained by a vertically extended corona simplified using two lamp-post sources \citep{Hancock2023}, and the corona itself may also vary within each individual observation \citep{Chainakun2023, Wilkins2023, Nakhonthong2024}. In these aspects, the model can be improved by being trained to reproduce the combined realistic response functions from different reprocessors or different geometry assumptions. It is also worth training the model using arbitrary shapes of the response functions in order to test how the reconstructed responses deviate from what is expected from the lamp-post reverberation.

While our focused information is limited to the coronal height, this work shows a potential way to extract the shape of the disc response function directly from the light curves. It is applicable to estimate the shape of the reverberation response in AGN within the scope of the lamp-post assumption. Particularly, we note that the complex features around the response peak still cannot be well-resolved with the current version of the model. This can be improved, in the future, in several ways such as by providing more synthetic data for training, but this undoubtedly requires a large amount of computational time. Reconstructing the reverberation response functions simultaneously with the responses due to other processes is vital for testing the models and concluding a unique interpretation of the source geometry. Implementing another type of neural network, such as convolutional neural network (CNN) and recurrent neural network (RNN), is another possibility to improve the prediction capability, flexibility, and sensitivity. A more comprehensive model that incorporates the warm corona is necessary for a fully understanding of different emission regions in the accretion flow and their contributions to the observed spectral and timing properties of AGN.

\section{Acknowledgements}
This work was supported by (i) Suranaree University of Technology (SUT), (ii) Thailand Science Research and Innovation (TSRI), and (iii) National Science, Research and Innovation Fund (grant number 204265). J.J. acknowledges support from Leverhulme Trust, Isaac Newton Trust and St Edmund's College, University of Cambridge. F.P.N. gratefully acknowledges the generous and invaluable support of the Klaus Tschira Foundation. F.P.N. acknowledges funding from the European Research Council (ERC) under the European Union's Horizon 2020 research and innovation program (grant agreement No 951549).

\bibliography{Ref}{}

\begin{thebibliography}{}
\expandafter\ifx\csname natexlab\endcsname\relax\def\natexlab#1{#1}\fi
\providecommand{\url}[1]{\href{#1}{#1}}
\providecommand{\dodoi}[1]{doi:~\href{http://doi.org/#1}{\nolinkurl{#1}}}
\providecommand{\doeprint}[1]{\href{http://ascl.net/#1}{\nolinkurl{http://ascl.net/#1}}}
\providecommand{\doarXiv}[1]{\href{https://arxiv.org/abs/#1}{\nolinkurl{https://arxiv.org/abs/#1}}}

\bibitem[{{Alston} {et~al.}(2014){Alston}, {Done}, \& {Vaughan}}]{Alston2014}
{Alston}, W.~N., {Done}, C., \& {Vaughan}, S. 2014, \mnras, 439, 1548, \dodoi{10.1093/mnras/stu005}

\bibitem[{Alston {et~al.}(2020)Alston, Fabian, Kara, Parker, Dovciak, Pinto, Jiang, Middleton, Miniutti, Walton, Wilkins, Buisson, Caballero-Garcia, Cackett, De~Marco, Gallo, Lohfink, Reynolds, Uttley, Young, \& Zogbhi}]{Alston2020}
Alston, W.~N., Fabian, A.~C., Kara, E., {et~al.} 2020, Nature Astronomy, 4, 597, \dodoi{10.1038/s41550-019-1002-x}

\bibitem[{{Ar{\'e}valo} \& {Uttley}(2006)}]{Arevalo2006}
{Ar{\'e}valo}, P., \& {Uttley}, P. 2006, \mnras, 367, 801, \dodoi{10.1111/j.1365-2966.2006.09989.x}

\bibitem[{{Ballantyne} {et~al.}(2024){Ballantyne}, {Sudhakar}, {Fairfax}, {Bianchi}, {Czerny}, {De Rosa}, {De Marco}, {Middei}, {Palit}, {Petrucci}, {R{\'o}{\.z}a{\'n}ska}, \& {Ursini}}]{Ballantyne2024}
{Ballantyne}, D.~R., {Sudhakar}, V., {Fairfax}, D., {et~al.} 2024, \mnras, 530, 1603, \dodoi{10.1093/mnras/stae944}

\bibitem[{{Caballero-Garc{\'\i}a} {et~al.}(2018){Caballero-Garc{\'\i}a}, {Papadakis}, {Dov{\v{c}}iak}, {Bursa}, {Epitropakis}, {Karas}, \& {Svoboda}}]{Caballero2018}
{Caballero-Garc{\'\i}a}, M.~D., {Papadakis}, I.~E., {Dov{\v{c}}iak}, M., {et~al.} 2018, \mnras, 480, 2650, \dodoi{10.1093/mnras/sty1990}

\bibitem[{{Caballero-Garc{\'\i}a} {et~al.}(2020){Caballero-Garc{\'\i}a}, {Papadakis}, {Dov{\v{c}}iak}, {Bursa}, {Svoboda}, \& {Karas}}]{Caballero2020}
---. 2020, \mnras, 498, 3184, \dodoi{10.1093/mnras/staa2554}

\bibitem[{{Cackett} {et~al.}(2021){Cackett}, {Bentz}, \& {Kara}}]{Cackett2021}
{Cackett}, E.~M., {Bentz}, M.~C., \& {Kara}, E. 2021, iScience, 24, 102557, \dodoi{10.1016/j.isci.2021.102557}

\bibitem[{{Cackett} {et~al.}(2014){Cackett}, {Zoghbi}, {Reynolds}, {Fabian}, {Kara}, {Uttley}, \& {Wilkins}}]{Cackett2014}
{Cackett}, E.~M., {Zoghbi}, A., {Reynolds}, C., {et~al.} 2014, \mnras, 438, 2980, \dodoi{10.1093/mnras/stt2424}

\bibitem[{{Carter} \& {Read}(2007)}]{Carter2007}
{Carter}, J.~A., \& {Read}, A.~M. 2007, \aap, 464, 1155, \dodoi{10.1051/0004-6361:20065882}

\bibitem[{{Chainakun}(2019)}]{Chainakun2019a}
{Chainakun}, P. 2019, \apj, 878, 20, \dodoi{10.3847/1538-4357/ab1f0a}

\bibitem[{{Chainakun} {et~al.}(2022){Chainakun}, {Luangtip}, {Jiang}, \& {Young}}]{Chainakun2022b}
{Chainakun}, P., {Luangtip}, W., {Jiang}, J., \& {Young}, A.~J. 2022, \apj, 934, 166, \dodoi{10.3847/1538-4357/ac7d55}

\bibitem[{{Chainakun} {et~al.}(2023){Chainakun}, {Nakhonthong}, {Luangtip}, \& {Young}}]{Chainakun2023}
{Chainakun}, P., {Nakhonthong}, N., {Luangtip}, W., \& {Young}, A.~J. 2023, \mnras, 523, 111, \dodoi{10.1093/mnras/stad1416}

\bibitem[{{Chainakun} \& {Young}(2015)}]{Chainakun2015}
{Chainakun}, P., \& {Young}, A.~J. 2015, \mnras, 452, 333, \dodoi{10.1093/mnras/stv1333}

\bibitem[{{Chainakun} {et~al.}(2016){Chainakun}, {Young}, \& {Kara}}]{Chainakun2016}
{Chainakun}, P., {Young}, A.~J., \& {Kara}, E. 2016, \mnras, 460, 3076, \dodoi{10.1093/mnras/stw1105}

\bibitem[{{Chelouche} {et~al.}(2019){Chelouche}, {Pozo Nu{\~n}ez}, \& {Kaspi}}]{Chelouche2019}
{Chelouche}, D., {Pozo Nu{\~n}ez}, F., \& {Kaspi}, S. 2019, Nature Astronomy, 3, 251, \dodoi{10.1038/s41550-018-0659-x}

\bibitem[{Cheng {et~al.}(2020)Cheng, Li, Conselice, Aragón-Salamanca, Dye, \& Metcalf}]{cheng_identifying_2020}
Cheng, T.-Y., Li, N., Conselice, C.~J., {et~al.} 2020, Monthly Notices of the Royal Astronomical Society, 494, 3750, \dodoi{10.1093/mnras/staa1015}

\bibitem[{{Chiang} {et~al.}(2015){Chiang}, {Walton}, {Fabian}, {Wilkins}, \& {Gallo}}]{Chiang2015}
{Chiang}, C.-Y., {Walton}, D.~J., {Fabian}, A.~C., {Wilkins}, D.~R., \& {Gallo}, L.~C. 2015, \mnras, 446, 759, \dodoi{10.1093/mnras/stu2087}

\bibitem[{{De Marco} {et~al.}(2013){De Marco}, {Ponti}, {Cappi}, {Dadina}, {Uttley}, {Cackett}, {Fabian}, \& {Miniutti}}]{DeMarco2013}
{De Marco}, B., {Ponti}, G., {Cappi}, M., {et~al.} 2013, \mnras, 431, 2441, \dodoi{10.1093/mnras/stt339}

\bibitem[{{Dov{\v{c}}iak} {et~al.}(2004{\natexlab{a}}){Dov{\v{c}}iak}, {Karas}, {Martocchia}, {Matt}, \& {Yaqoob}}]{Dovciak2004b}
{Dov{\v{c}}iak}, M., {Karas}, V., {Martocchia}, A., {Matt}, G., \& {Yaqoob}, T. 2004{\natexlab{a}}, in RAGtime 4/5: Workshops on black holes and neutron stars, 33--73, \dodoi{10.48550/arXiv.astro-ph/0407330}

\bibitem[{{Dov{\v{c}}iak} {et~al.}(2004{\natexlab{b}}){Dov{\v{c}}iak}, {Karas}, \& {Yaqoob}}]{Dovciak2004a}
{Dov{\v{c}}iak}, M., {Karas}, V., \& {Yaqoob}, T. 2004{\natexlab{b}}, \apjs, 153, 205, \dodoi{10.1086/421115}

\bibitem[{{Emmanoulopoulos} {et~al.}(2014){Emmanoulopoulos}, {Papadakis}, {Dov{\v c}iak}, \& {McHardy}}]{Emmanoulopoulos2014}
{Emmanoulopoulos}, D., {Papadakis}, I.~E., {Dov{\v c}iak}, M., \& {McHardy}, I.~M. 2014, \mnras, 439, 3931, \dodoi{10.1093/mnras/stu249}

\bibitem[{{Emmanoulopoulos} {et~al.}(2016){Emmanoulopoulos}, {Papadakis}, {Epitropakis}, {Pech{\'a}{\v{c}}ek}, {Dov{\v{c}}iak}, \& {McHardy}}]{Emmanoulopoulos2016}
{Emmanoulopoulos}, D., {Papadakis}, I.~E., {Epitropakis}, A., {et~al.} 2016, \mnras, 461, 1642, \dodoi{10.1093/mnras/stw1359}

\bibitem[{{Epitropakis} {et~al.}(2016){Epitropakis}, {Papadakis}, {Dov{\v{c}}iak}, {Pech{\'a}{\v{c}}ek}, {Emmanoulopoulos}, {Karas}, \& {McHardy}}]{Epitropakis2016}
{Epitropakis}, A., {Papadakis}, I.~E., {Dov{\v{c}}iak}, M., {et~al.} 2016, \aap, 594, A71, \dodoi{10.1051/0004-6361/201527748}

\bibitem[{{Fabian} {et~al.}(2009){Fabian}, {Zoghbi}, {Ross}, {Uttley}, {Gallo}, {Brandt}, {Blustin}, {Boller}, {Caballero-Garcia}, {Larsson}, {Miller}, {Miniutti}, {Ponti}, {Reis}, {Reynolds}, {Tanaka}, \& {Young}}]{Fabian2009}
{Fabian}, A.~C., {Zoghbi}, A., {Ross}, R.~R., {et~al.} 2009, \nat, 459, 540, \dodoi{10.1038/nature08007}

\bibitem[{Fritsch \& Butland(1984)}]{PCHIP}
Fritsch, F.~N., \& Butland, J. 1984, SIAM Journal on Scientific and Statistical Computing, 5, 300, \dodoi{10.1137/0905021}

\bibitem[{{Frontera-Pons, J.} {et~al.}(2017){Frontera-Pons, J.}, {Sureau, F.}, {Bobin, J.}, \& {Le Floc’h, E.}}]{frontera2017}
{Frontera-Pons, J.}, {Sureau, F.}, {Bobin, J.}, \& {Le Floc’h, E.} 2017, A\&A, 603, A60, \dodoi{10.1051/0004-6361/201630240}

\bibitem[{{Garc{\'\i}a} {et~al.}(2013){Garc{\'\i}a}, {Dauser}, {Reynolds}, {Kallman}, {McClintock}, {Wilms}, \& {Eikmann}}]{Garcia2013}
{Garc{\'\i}a}, J., {Dauser}, T., {Reynolds}, C.~S., {et~al.} 2013, \apj, 768, 146, \dodoi{10.1088/0004-637X/768/2/146}

\bibitem[{{Garc{\'\i}a} \& {Kallman}(2010)}]{Garcia2010}
{Garc{\'\i}a}, J., \& {Kallman}, T.~R. 2010, \apj, 718, 695, \dodoi{10.1088/0004-637X/718/2/695}

\bibitem[{{George} \& {Fabian}(1991)}]{George1991}
{George}, I.~M., \& {Fabian}, A.~C. 1991, \mnras, 249, 352, \dodoi{10.1093/mnras/249.2.352}

\bibitem[{{Gonz{\'a}lez-Mart{\'\i}n} \& {Vaughan}(2012)}]{Gonzalez2012}
{Gonz{\'a}lez-Mart{\'\i}n}, O., \& {Vaughan}, S. 2012, \aap, 544, A80, \dodoi{10.1051/0004-6361/201219008}

\bibitem[{Goodfellow {et~al.}(2016)Goodfellow, Bengio, \& Courville}]{Goodfellow-et-al-2016}
Goodfellow, I., Bengio, Y., \& Courville, A. 2016, Deep Learning (MIT Press)

\bibitem[{{Hancock} {et~al.}(2023){Hancock}, {Young}, \& {Chainakun}}]{Hancock2023}
{Hancock}, S., {Young}, A.~J., \& {Chainakun}, P. 2023, \mnras, 520, 180, \dodoi{10.1093/mnras/stad144}

\bibitem[{{Huppenkothen} {et~al.}(2019){Huppenkothen}, {Bachetti}, {Stevens}, {Migliari}, {Balm}, {Hammad}, {Khan}, {Mishra}, {Rashid}, {Sharma}, {Martinez Ribeiro}, \& {Valles Blanco}}]{Huppenkothen2019}
{Huppenkothen}, D., {Bachetti}, M., {Stevens}, A.~L., {et~al.} 2019, \apj, 881, 39, \dodoi{10.3847/1538-4357/ab258d}

\bibitem[{{Jaiswal} {et~al.}(2023){Jaiswal}, {Prince}, {Panda}, \& {Czerny}}]{Jaiswal2023}
{Jaiswal}, V.~K., {Prince}, R., {Panda}, S., \& {Czerny}, B. 2023, \aap, 670, A147, \dodoi{10.1051/0004-6361/202244352}

\bibitem[{{Jansen} {et~al.}(2001){Jansen}, {Lumb}, {Altieri}, {Clavel}, {Ehle}, {Erd}, {Gabriel}, {Guainazzi}, {Gondoin}, {Much}, {Munoz}, {Santos}, {Schartel}, {Texier}, \& {Vacanti}}]{Jansen2001}
{Jansen}, F., {Lumb}, D., {Altieri}, B., {et~al.} 2001, \aap, 365, L1, \dodoi{10.1051/0004-6361:20000036}

\bibitem[{{Jiang} {et~al.}(2022){Jiang}, {Dauser}, {Fabian}, {Alston}, {Gallo}, {Parker}, \& {Reynolds}}]{Jiang2022}
{Jiang}, J., {Dauser}, T., {Fabian}, A.~C., {et~al.} 2022, \mnras, 514, 1107, \dodoi{10.1093/mnras/stac1144}

\bibitem[{{Jiang} {et~al.}(2018){Jiang}, {Parker}, {Fabian}, {Alston}, {Buisson}, {Cackett}, {Chiang}, {Dauser}, {Gallo}, {Garc{\'\i}a}, {Harrison}, {Lohfink}, {De Marco}, {Kara}, {Miller}, {Miniutti}, {Pinto}, {Walton}, \& {Wilkins}}]{Jiang2018}
{Jiang}, J., {Parker}, M.~L., {Fabian}, A.~C., {et~al.} 2018, \mnras, 477, 3711, \dodoi{10.1093/mnras/sty836}

\bibitem[{{Kallman} \& {Bautista}(2001)}]{Kallman2001}
{Kallman}, T., \& {Bautista}, M. 2001, \apjs, 133, 221, \dodoi{10.1086/319184}

\bibitem[{Kang(2021)}]{Kang_2021}
Kang, M.~J. 2021, Pytorch-VAE-tutorial: A simple tutorial of Variational AutoEncoders with Pytorch.
\newblock \url{https://github.com/Jackson-Kang/Pytorch-VAE-tutorial/tree/master}

\bibitem[{{Kara} {et~al.}(2016){Kara}, {Alston}, {Fabian}, {Cackett}, {Uttley}, {Reynolds}, \& {Zoghbi}}]{Kara2016}
{Kara}, E., {Alston}, W.~N., {Fabian}, A.~C., {et~al.} 2016, \mnras, 462, 511, \dodoi{10.1093/mnras/stw1695}

\bibitem[{{Khanthasombat} {et~al.}(2024){Khanthasombat}, {Chainakun}, \& {Young}}]{Khanthasombat2024}
{Khanthasombat}, K., {Chainakun}, P., \& {Young}, A.~J. 2024, \mnras, 528, 3130, \dodoi{10.1093/mnras/stae173}

\bibitem[{{Kotov} {et~al.}(2001){Kotov}, {Churazov}, \& {Gilfanov}}]{Kotov2001}
{Kotov}, O., {Churazov}, E., \& {Gilfanov}, M. 2001, \mnras, 327, 799, \dodoi{10.1046/j.1365-8711.2001.04769.x}

\bibitem[{{Kubota} \& {Done}(2018)}]{Kubota2018}
{Kubota}, A., \& {Done}, C. 2018, \mnras, 480, 1247, \dodoi{10.1093/mnras/sty1890}

\bibitem[{{Kumari} {et~al.}(2024){Kumari}, {Dewangan}, {Papadakis}, \& {Singh}}]{Kumari2024}
{Kumari}, K., {Dewangan}, G.~C., {Papadakis}, I.~E., \& {Singh}, K.~P. 2024, \mnras, 527, 5668, \dodoi{10.1093/mnras/stad3444}

\bibitem[{{Kumari} {et~al.}(2023){Kumari}, {Jana}, {Naik}, \& {Nandi}}]{Kumari2023}
{Kumari}, N., {Jana}, A., {Naik}, S., \& {Nandi}, P. 2023, \mnras, 521, 5440, \dodoi{10.1093/mnras/stad867}

\bibitem[{{Lyubarskii}(1997)}]{Lyubarskii1997}
{Lyubarskii}, Y.~E. 1997, \mnras, 292, 679, \dodoi{10.1093/mnras/292.3.679}

\bibitem[{{Mankatwit} {et~al.}(2023){Mankatwit}, {Chainakun}, {Luangtip}, \& {Young}}]{Mankatwit2023}
{Mankatwit}, N., {Chainakun}, P., {Luangtip}, W., \& {Young}, A.~J. 2023, \mnras, 523, 4080, \dodoi{10.1093/mnras/stad1706}

\bibitem[{{McHardy} {et~al.}(2007){McHardy}, {Ar{\'e}valo}, {Uttley}, {Papadakis}, {Summons}, {Brinkmann}, \& {Page}}]{McHardy2007}
{McHardy}, I.~M., {Ar{\'e}valo}, P., {Uttley}, P., {et~al.} 2007, \mnras, 382, 985, \dodoi{10.1111/j.1365-2966.2007.12411.x}

\bibitem[{{Middei} {et~al.}(2018){Middei}, {Bianchi}, {Cappi}, {Petrucci}, {Ursini}, {Arav}, {Behar}, {Branduardi-Raymont}, {Costantini}, {De Marco}, {Di Gesu}, {Ebrero}, {Kaastra}, {Kaspi}, {Kriss}, {Mao}, {Mehdipour}, {Paltani}, {Peretz}, \& {Ponti}}]{Middei2018}
{Middei}, R., {Bianchi}, S., {Cappi}, M., {et~al.} 2018, \aap, 615, A163, \dodoi{10.1051/0004-6361/201832726}

\bibitem[{{Miniutti} \& {Fabian}(2004)}]{Miniutti2004}
{Miniutti}, G., \& {Fabian}, A.~C. 2004, \mnras, 349, 1435, \dodoi{10.1111/j.1365-2966.2004.07611.x}

\bibitem[{{Nakhonthong} {et~al.}(2024){Nakhonthong}, {Chainakun}, {Luangtip}, \& {Young}}]{Nakhonthong2024}
{Nakhonthong}, N., {Chainakun}, P., {Luangtip}, W., \& {Young}, A.~J. 2024, arXiv e-prints, arXiv:2404.04493, \dodoi{10.48550/arXiv.2404.04493}

\bibitem[{Orwat-Kapola {et~al.}(2021)Orwat-Kapola, Bird, Hill, Altamirano, \& Huppenkothen}]{OrwatKapola2021}
Orwat-Kapola, J.~K., Bird, A.~J., Hill, A.~B., Altamirano, D., \& Huppenkothen, D. 2021, Monthly Notices of the Royal Astronomical Society, 509, 1269, \dodoi{10.1093/mnras/stab3043}

\bibitem[{{Papadakis} {et~al.}(2016){Papadakis}, {Pech{\'a}{\v{c}}ek}, {Dov{\v{c}}iak}, {Epitropakis}, {Emmanoulopoulos}, \& {Karas}}]{Papadakis2016}
{Papadakis}, I., {Pech{\'a}{\v{c}}ek}, T., {Dov{\v{c}}iak}, M., {et~al.} 2016, \aap, 588, A13, \dodoi{10.1051/0004-6361/201527246}

\bibitem[{Paszke {et~al.}(2019)Paszke, Gross, Massa, Lerer, Bradbury, Chanan, Killeen, Lin, Gimelshein, Antiga, Desmaison, Köpf, Yang, DeVito, Raison, Tejani, Chilamkurthy, Steiner, Fang, Bai, \& Chintala}]{paszke2019pytorch}
Paszke, A., Gross, S., Massa, F., {et~al.} 2019, PyTorch: An Imperative Style, High-Performance Deep Learning Library.
\newblock \doarXiv{1912.01703}

\bibitem[{Peterson {et~al.}(2004)Peterson, Ferrarese, Gilbert, Kaspi, Malkan, Maoz, Merritt, Netzer, Onken, Pogge, Vestergaard, \& Wandel}]{Peterson_2004}
Peterson, B.~M., Ferrarese, L., Gilbert, K.~M., {et~al.} 2004, The Astrophysical Journal, 613, 682, \dodoi{10.1086/423269}

\bibitem[{{Porquet} {et~al.}(2018){Porquet}, {Reeves}, {Matt}, {Marinucci}, {Nardini}, {Braito}, {Lobban}, {Ballantyne}, {Boggs}, {Christensen}, {Dauser}, {Farrah}, {Garcia}, {Hailey}, {Harrison}, {Stern}, {Tortosa}, {Ursini}, \& {Zhang}}]{Porquet2018}
{Porquet}, D., {Reeves}, J.~N., {Matt}, G., {et~al.} 2018, \aap, 609, A42, \dodoi{10.1051/0004-6361/201731290}

\bibitem[{{Pozo Nu{\~n}ez} {et~al.}(2023){Pozo Nu{\~n}ez}, {Bruckmann}, {Deesamutara}, {Czerny}, {Panda}, {Lobban}, {Pietrzy{\'n}ski}, \& {Polsterer}}]{Nunez2023}
{Pozo Nu{\~n}ez}, F., {Bruckmann}, C., {Deesamutara}, S., {et~al.} 2023, \mnras, 522, 2002, \dodoi{10.1093/mnras/stad286}

\bibitem[{Pozo~Nuñez {et~al.}(2023)Pozo~Nuñez, Gianniotis, \& Polsterer}]{nunez_gaussian_2023}
Pozo~Nuñez, F., Gianniotis, N., \& Polsterer, K.~L. 2023, Astronomy \& Astrophysics, 674, A83, \dodoi{10.1051/0004-6361/202345932}

\bibitem[{{Reynolds} \& {Nowak}(2003)}]{Reynolds2003}
{Reynolds}, C.~S., \& {Nowak}, M.~A. 2003, \physrep, 377, 389, \dodoi{10.1016/S0370-1573(02)00584-7}

\bibitem[{{Reynolds} {et~al.}(1999){Reynolds}, {Young}, {Begelman}, \& {Fabian}}]{Reynolds1999}
{Reynolds}, C.~S., {Young}, A.~J., {Begelman}, M.~C., \& {Fabian}, A.~C. 1999, \apj, 514, 164, \dodoi{10.1086/306913}

\bibitem[{{Ross} \& {Fabian}(2005)}]{Ross2005}
{Ross}, R.~R., \& {Fabian}, A.~C. 2005, \mnras, 358, 211, \dodoi{10.1111/j.1365-2966.2005.08797.x}

\bibitem[{{Ross} {et~al.}(1999){Ross}, {Fabian}, \& {Young}}]{Ross1999}
{Ross}, R.~R., {Fabian}, A.~C., \& {Young}, A.~J. 1999, \mnras, 306, 461, \dodoi{10.1046/j.1365-8711.1999.02528.x}

\bibitem[{Sacchi(2016)}]{SNR_generator}
Sacchi, M.~D. 2016, Adding noise with a desired signal-to-noise ratio, \url{https://sites.ualberta.ca/~msacchi/SNR_Def.pdf}

\bibitem[{{Ursini} {et~al.}(2020){Ursini}, {Petrucci}, {Bianchi}, {Matt}, {Middei}, {Marcel}, {Ferreira}, {Cappi}, {De Marco}, {De Rosa}, {Malzac}, {Marinucci}, {Ponti}, \& {Tortosa}}]{Ursini2020}
{Ursini}, F., {Petrucci}, P.~O., {Bianchi}, S., {et~al.} 2020, \aap, 634, A92, \dodoi{10.1051/0004-6361/201936486}

\bibitem[{{Uttley} {et~al.}(2014){Uttley}, {Cackett}, {Fabian}, {Kara}, \& {Wilkins}}]{Uttley2014}
{Uttley}, P., {Cackett}, E.~M., {Fabian}, A.~C., {Kara}, E., \& {Wilkins}, D.~R. 2014, \aapr, 22, 72, \dodoi{10.1007/s00159-014-0072-0}

\bibitem[{{Vaughan} {et~al.}(2003){Vaughan}, {Edelson}, {Warwick}, \& {Uttley}}]{Vaughan2003}
{Vaughan}, S., {Edelson}, R., {Warwick}, R.~S., \& {Uttley}, P. 2003, \mnras, 345, 1271, \dodoi{10.1046/j.1365-2966.2003.07042.x}

\bibitem[{{V{\'e}ron-Cetty} \& {V{\'e}ron}(2006)}]{Veron-Cetty2006}
{V{\'e}ron-Cetty}, M.~P., \& {V{\'e}ron}, P. 2006, \aap, 455, 773, \dodoi{10.1051/0004-6361:20065177}

\bibitem[{{Wilkins}(2023)}]{Wilkins2023}
{Wilkins}, D.~R. 2023, \mnras, 526, 3441, \dodoi{10.1093/mnras/stad2936}

\bibitem[{{Wilkins} \& {Fabian}(2013)}]{Wilkins2013}
{Wilkins}, D.~R., \& {Fabian}, A.~C. 2013, \mnras, 430, 247, \dodoi{10.1093/mnras/sts591}

\bibitem[{{Wilkins} \& {Gallo}(2015)}]{Wilkins2015}
{Wilkins}, D.~R., \& {Gallo}, L.~C. 2015, \mnras, 449, 129, \dodoi{10.1093/mnras/stv162}

\bibitem[{{Xu} {et~al.}(2021){Xu}, {Ding}, {Gu}, {Guo}, \& {Contini}}]{Xu2021}
{Xu}, X., {Ding}, N., {Gu}, Q., {Guo}, X., \& {Contini}, E. 2021, \mnras, 507, 3572, \dodoi{10.1093/mnras/stab2278}

\bibitem[{{Young} \& {Reynolds}(2000)}]{Young2000}
{Young}, A.~J., \& {Reynolds}, C.~S. 2000, \apj, 529, 101, \dodoi{10.1086/308236}

\bibitem[{{Zoghbi} {et~al.}(2010){Zoghbi}, {Fabian}, {Uttley}, {Miniutti}, {Gallo}, {Reynolds}, {Miller}, \& {Ponti}}]{Zoghbi2010}
{Zoghbi}, A., {Fabian}, A.~C., {Uttley}, P., {et~al.} 2010, \mnras, 401, 2419, \dodoi{10.1111/j.1365-2966.2009.15816.x}

\end{thebibliography}
\bibliographystyle{aasjournal}

\appendix
\section{Signal-to-Noise Ratio} \label{app:SNR}
In this work, SNR is defined by the fraction between signal and noise power spectra, assuming noise involvement in all frequency domain, as mentioned in Equation~\ref{eq:s/n}. We also note that while the SNR obtained by traditional method and the method in this paper may not directly correlate, somehow, they still have the correlation at some level. Indeed, the observations that have high SNR measured by traditional method (e.g., Table 1  of \citealt{Nakhonthong2024}) are likely to have a relatively high SNR when it is calculated by our method (e.g., Obs. ID 0780561601, 0792180401 and 0792180601). Thus, we demonstrate how the behavior of light curves in the simulated, training REV+PROP+NG+NP data varies with SNR. For comparison, we present the light curve of IRAS~13224--3809 (obs. ID 067358401) showing SNR = 62.23, which is almost similar in quality to our example light curve with a similar SNR.
\begin{figure}[h!]
    \centering
    \includegraphics[width=0.99\linewidth]{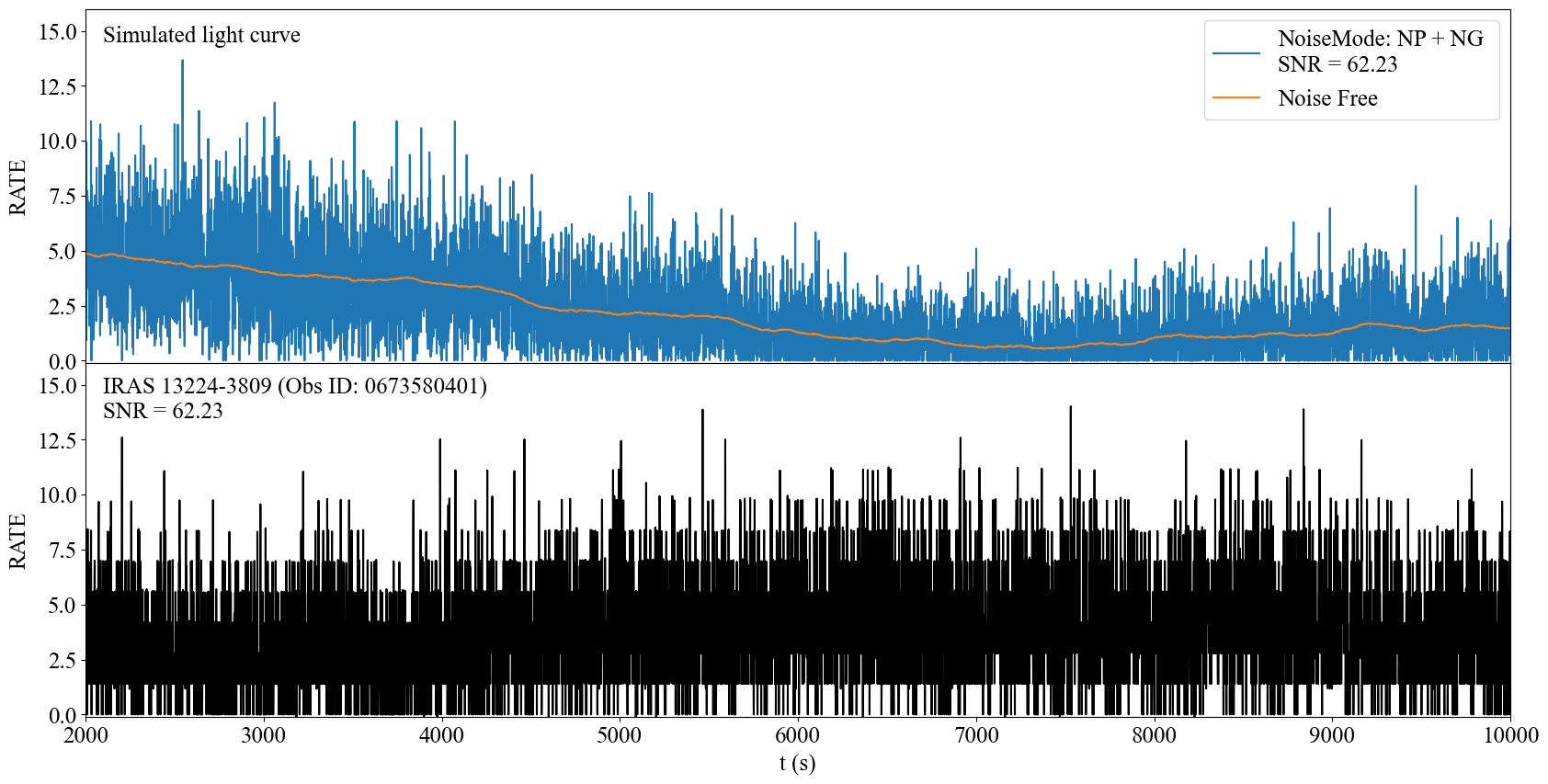}
    \caption{Top panel: Simulated light curves. Noise-free and noisy light curves are labeled as orange and blue lines, respectively. Bottom panel: Example of IRAS~13224-3809 light curve showing SNR = 62.23.}
    \label{fig:snr_scenarios}
\end{figure}

\section{Investigation of parameter variation effect to prediction efficiency} \label{app:perfeval}
Furthering our performance investigation, we test for the general applicability of the model when the observed data have slightly different characteristics to the training data. Firstly, we investigate case when the reflection fraction $R$ is outside the range using to train the model. Note that, in this investigation, the model is trained with $R \in [1,3]$, while the $R$ range is extended to $[1,10]$ in the final model applied to IRAS~13224--3809. We also produce some new response functions with different $\Gamma$, $i$, and $M$ within the observed uncertainty reported in \cite{Alston2020}\footnote{Varying parameters are $2 \leq \Gamma \leq 2.68$, $62^{\circ} \leq i \leq 72^{\circ}$, and $1.7 \leq M~(10^{6}~M_{\odot}) \leq 2.1$}. We generate sets of convolved light curves corresponding to four scenarios including 1) only $R$ is varied $\sim 25\%$ beyond the trained values, 2) only $\Gamma$ is varied, 3) $\Gamma$ and $i$ are varied, and 4) $\Gamma$, $i$ and $M$ are varied. We test the original REV+PROP model (trained using the light curves with the fixed $\Gamma$, $i$ and $M$) on these light curves. The results are reported in Table ~\ref{tab:varyParameters}. We find that the model can capture the essential aspects of the coronal height while accommodating potential variations in the reflection fraction. This is likely because, although the overall shape and amplitude of the response function vary, the first response time of the reverberation delays--which is closely related to the coronal height--remains unaffected. The model appears to use this initial response signature as a key indicator for determining the coronal height. While the response functions extracted by the model are still generally excellence in terms of signal coherence, we can see a decrease in the cross-correlation. This is expected since the training data have slightly different characteristics to the observed data, even though this should be the real scenario where we do not exactly know the precise values of these parameters. Regardless, this $R$ range is likely adequate for applying the model to the specific AGN IRAS~13224--3809, with cross-correlation $\gtrsim 0.8$ and signal coherence $\gtrsim 0.9$. Finally, it could be expected that the performance shall decline when Gaussian and Poisson noise are considered, in the similar trend as shown in Figure~\ref{fig:sc_noise}.

\begin{table}[h!]
    \centering
    \caption{Prediction performance of the original REV+PROP model (trained with the light curves where $\Gamma$, $i$ and $M$ are fixed) on the test data produced in four scenarios.}
    \begin{tabular}{|l|c|c|}
         \hline
         \multicolumn{1}{|c|}{Scenario} &  Cross-Correlation & Signal Coherence\\
         \hline
         vary $R$ outside the trained range &  $0.939^{+0.047}_{-0.119}$ & $0.964^{+0.016}_{-0.088}$\\
         \hline
         vary $\Gamma$ &  $0.875^{+0.055}_{-0.095}$ & $0.959^{+0.005}_{-0.078}$\\
         \hline
         vary $\Gamma$ and $i$ &  $0.861^{+0.080}_{-0.156}$ & $0.959^{+0.009}_{-0.025}$\\
         \hline
         vary $\Gamma$, $i$, and $M$ &  $0.848^{+0.116}_{-0.110}$ & $0.959^{+0.007}_{-0.025}$\\
         \hline
    \end{tabular}
    \label{tab:varyParameters}
\end{table}

\end{document}